\documentclass[10pt]{article}
\usepackage{graphicx,float,latexsym,amssymb,subeqnarray,colordvi}
\usepackage[latin1]{inputenc}
\usepackage{a4wide,amsmath,amsfonts,amssymb,subeqnarray}

\def\R{\mathbb{R}}
\def\Z{\mathbb{Z}}

\def\imi{\textbf{\hskip1pt i\hskip1pt}}

\newcommand{\lleft}{\rm{left}}
\newcommand{\rright}{\rm{right}}
\newcommand{\iin}{\rm{int}}

\begin{document}
\title{ADI finite difference schemes\\for the Heston--Hull--White PDE}
\author{Tinne~Haentjens\footnote{Department of Mathematics and Computer Science,
University of Antwerp, Middelheimlaan 1, B-2020 Antwerp, Belgium.
\mbox{Email}: \texttt{\{tinne.haentjens,karel.inthout\}@ua.ac.be}.}
~and Karel~J.~in 't Hout\footnotemark[\value{footnote}]
}
\date{\today}

\maketitle
\numberwithin{equation}{section}

\begin{abstract}
In this paper we investigate the effectiveness of Alternating Direction
Implicit (ADI) time discretization schemes in the numerical solution of
the three-dimensional Heston--Hull--White partial differential equation,
which is semidiscretized by applying finite difference schemes on
nonuniform spatial grids.
We consider the Heston--Hull--White model with arbitrary correlation factors,
with time-dependent mean-reversion levels, with short and long maturities,
for cases where the Feller condition is satisfied and for cases where it is
not.
In addition, both European-style call options and up-and-out call options 
are considered.
It is shown through extensive tests that ADI schemes, with a proper choice
of their parameters, perform very well in all situations - in terms of
stability, accuracy and efficiency.
\end{abstract}
\vspace{5mm}

\setcounter{equation}{0}
\section{Introduction}
The main aim of this paper is to investigate the effectiveness of
Alternating Direction Implicit (ADI) time discretization schemes in
the numerical solution of three-dimensional time-dependent partial
differential equations (PDEs) arising in financial option valuation
theory.
As a prototype case we consider the Heston--Hull--White PDE, but our
conclusions concerning ADI schemes extend to many other related
three-dimensional models.

Consider the asset price process given by the system of stochastic
differential equations (SDEs)
\begin{equation}\label{SDE}
\begin{cases}
&d S_\tau\, = R_\tau S_\tau\, d\tau + \sqrt{V_\tau}\, S_\tau\, d W^1_\tau\,,\\
&d V_\tau\, = \kappa(\eta -V_\tau)\, d\tau + \sigma_1\sqrt{V_\tau}\, d W^2_\tau\,,\\
&d R_\tau = a(b(\tau) -R_\tau)\, d\tau + \sigma_2\, d W^3_\tau\,.
\end{cases}
\end{equation}
The random variables $S_\tau$, $V_\tau$, $R_\tau$ represent, respectively,
the asset price, its variance and the interest rate at time $\tau>0$.
The parameters $\kappa$, $\eta$, $\sigma_1$ and $a$, $\sigma_2$ are given
positive real constants and $b$ denotes a given deterministic, positive
function of time, called the mean-reversion level.
The $W^1_\tau$, $W^2_\tau$, $W^3_\tau$ are Brownian motions under a
risk-neutral measure with given correlation factors $\rho_{12}$,
$\rho_{13}$, $\rho_{23}\in [-1,1]$ such that the pertinent correlation
matrix is positive semidefinite.

The asset price model (\ref{SDE}) can be viewed as an extension of the
popular Heston stochastic volatility model (Heston (1993)) where the
interest rate is not constant but also follows a stochastic process,
described here by the Hull--White model (Hull \& White (1990)).
The function $b$ is chosen as to match the current term structure of
interest rates.
The hybrid Heston--Hull--White model (\ref{SDE}) has recently been
studied in Giese (2006), Muskulus, In 't Hout, Bierkens et al (2007),
Grzelak, Oosterlee \& Van~Weeren (2009), Grzelak \& Oosterlee (2011)
and can lead to a more accurate valuation of option products that
are sensitive to both volatility and interest rates.

Let $T>0$ be given.
If at time $\tau \in [0,T)$ the asset price equals $s$, the variance equals
$v$ and the interest rate equals $r$, then for a European-style option with
maturity time $T$ and payoff function $\phi$ the risk-neutral value is given
by
\begin{equation}\label{price}
\varphi(s,v,r,\tau)=
\mathbb{E}\left[e^{-\int_\tau^T R_{\varsigma}\,d\varsigma}\,\phi(S_T,V_T,R_T)
\bigm\vert S_\tau=s,\,V_\tau=v,\,R_\tau=r\right],
\end{equation}
where $\mathbb{E}$ denotes conditional expectation under the risk-neutral
measure.
In this paper we consider $u(s,v,r,t) = \varphi(s,v,r,T-t)$.
Common arguments in financial mathematics imply that if the option value
function $u$ is sufficiently smooth then it satisfies the PDE
\begin{align}\label{HHWPDE}
\frac{\partial u}{\partial t} =
&~\tfrac{1}{2}s^2v\frac{\partial^2 u}{\partial s^2}
+ \tfrac{1}{2}\sigma_1^2 v \frac{\partial^2 u}{\partial v^2}
+ \tfrac{1}{2}\sigma_2^2\frac{\partial^2 u}{\partial r^2}\nonumber\\
&~+ \rho_{12} \sigma_1 s v \frac{\partial^2 u}{\partial s \partial v}
+ \rho_{13} \sigma_2 s \sqrt{v} \frac{\partial^2 u}{\partial s \partial r}
+ \rho_{23} \sigma_1 \sigma_2 \sqrt{v} \frac{\partial^2 u}{\partial v \partial r}\nonumber\\
&~+ rs \frac{\partial u}{\partial s} + \kappa (\eta - v)\frac{\partial u}{\partial v}
+ a(b(T-t)-r)\frac{\partial u}{\partial r} - ru
\end{align}
for $s>0$, $v>0$, $-\infty < r < \infty$ and $0< t \le T$.
Here $-\infty < r < \infty$ since the Hull--White model yields any,
positive or negative, value for the interest rate.
We refer to (\ref{HHWPDE}) as the {\it Heston--Hull--White (HHW) PDE}.\,
It forms a time-dependent convection-diffusion-reaction equation on an
unbounded, three-dimensional spatial domain.
The HHW PDE contains three mixed spatial-derivative terms, stemming
from the correlations between the underlying Brownian motions.
Next, if $v\downarrow 0$ then all second-order derivative terms, apart
from the $\partial^2 u/ \partial r^2$ term, vanish.
This degeneracy feature is already familiar from other financial PDEs,
such as the Heston PDE.
Finally, we note that the coefficient of the $\partial u /\partial r$
term is time-dependent.

The HHW PDE is complemented by initial and boundary conditions
that are determined by the specific option under consideration.
The initial condition is given by the payoff function,
\begin{equation}\label{IC}
u(s,v,r,0) = \phi(s,v,r).
\end{equation}
Boundary conditions will be discussed below.

The initial-boundary value problem for the HHW PDE does not admit analytic
solutions in (semi) closed-form in general.
An exception concerns European call options if the two correlations
$\rho_{13}$ and $\rho_{23}$ are equal to zero.
Then a direct extension of Heston's (1993) formula is available; it is
given in the Appendix.

For the numerical solution of the HHW PDE we consider the well-known and
versatile method-of-lines approach, see eg, Hundsdorfer \& Verwer (2003).
Here the PDE is first discretized in the spatial variables $s$,~$v$,~$r$.
This leads to a system of stiff ordinary differential equations, the
so-called semidiscrete system, which is subsequently solved by applying
a suitable time discretization method.
Since the HHW PDE is three-dimensional, the obtained semidiscrete systems
are very large and also possess a large bandwidth.
As a consequence, the selection of the time discretization method
is critical for its effective numerical solution.
To this purpose, we analyze in the present paper splitting schemes
of the ADI type.

An outline of the rest of our paper is as follows.
In Section~\ref{space} we describe the spatial discretization of the
HHW PDE. Here finite difference schemes on nonuniform spatial grids
are applied.
In Section~\ref{ADI} we formulate and discuss the four ADI schemes
under consideration in this paper: the Douglas scheme, the Craig--Sneyd
scheme, the modified Craig--Sneyd scheme and the Hundsdorfer--Verwer
scheme.
In Section~\ref{numexp} extensive numerical tests with these ADI schemes
are presented.
Here we investigate in detail the temporal discretization errors.
Our tests include arbitrary correlation factors, time-dependent
mean-reversion levels, cases where the Feller condition is satisfied and
cases where it is not.
In addition, both European call options and up-and-out call options
are considered.
Section \ref{concl} gives conclusions and issues for future research.


\setcounter{equation}{0}
\section{Space discretization of the HHW PDE}\label{space}
In this section we describe the spatial discretization of the HHW PDE.
For ease of presentation, we consider here European call options.
Thus $\phi(s,v,r) = \max(0,s-K)$ with given strike price $K>0$.
The spatial discretization is readily adapted to various exotic
options; cf also Section~\ref{numexp}.

\subsection{Boundary conditions}\label{BCs}
For the semidiscretization, the spatial domain is first restricted
to a bounded set
$[0,S_{\max}]\times [0,V_{\max}] \times [-R_{\max},R_{\max}]$
with fixed values $S_{\max}$, $V_{\max}$, $R_{\max}$ chosen
sufficiently large.
The following boundary conditions are imposed,
\begin{subeqnarray}\label{BC}
\phantom{\frac{\partial u}{\partial s}}u(s,v,r,t)&=~~0
\quad &{\rm whenever}~~s=0,\\
\frac{\partial u}{\partial s}(s,v,r,t)&=~~1
\quad &{\rm whenever}~~s=S_{\max},\\
\phantom{\frac{\partial u}{\partial s}}u(s,v,r,t)&=~~s
\quad &{\rm whenever}~~v=V_{\max},\\
\frac{\partial u}{\partial r}(s,v,r,t)&=~~0
\quad &{\rm whenever}~~r=\pm R_{\max}.
\end{subeqnarray}
\vskip0.1cm\noindent
Clearly these conditions are of Dirichlet and Neumann type.
Condition (\ref{BC}a) is obvious, (\ref{BC}b) and (\ref{BC}c)
have already been used in the literature for the Heston PDE,
and (\ref{BC}d) appears to be new.
Concerning the latter condition, it is straightforward to prove that
under the Black--Scholes model the rho of a European call option
vanishes for extreme values of the spot interest rate, and it is
plausible that this holds under the asset price model (\ref{SDE}) as
well.

At the important, special boundary $v=0$ we consider inserting
$v=0$ into the HHW PDE.\footnote{We are grateful to Peter Forsyth
for a stimulating discussion on this issue.}
This is motivated by a theorem of Ekstr\"{o}m \& Tysk (2011)
revealing that in the Cox--Ingersoll--Ross model, which corresponds
to $V_\tau$ in the SDE (\ref{SDE}), the resulting equation is
fulfilled by the risk-neutral option value.
We note the remarkable fact that this holds irrespective of whether
or not the Feller condition $2\kappa\eta > \sigma_1^2$, well-known
from the SDE literature, is satisfied.

\subsection{Spatial grid}\label{grid}
The HHW PDE is semidiscretized on a nonuniform Cartesian spatial grid.
The nonuniform grid defined in this section is advantageous over a
uniform one. This will be illustrated by numerical experiments in
Section \ref{numexp}.

In the $s$-direction we consider placing relatively many mesh points
throughout a given interval $[S_{\lleft},S_{\rright}]\subset [0,S_{\max}]$
containing the strike $K$.
This is natural, firstly, because this is the region of interest in
applications, and secondly, it alleviates numerical difficulties due
to the initial (payoff) function $\phi$ that has a discontinuous
derivative at $s=K$.
Let integer $m_1 \ge 1$ and parameter $d_1>0$ and let equidistant
points
\mbox{$\xi_{\min}=\xi_0 < \xi_1 < \ldots < \xi_{m_1}=\xi_{\max}$}
be given with
\begin{align*}
\xi_{\min} &= \sinh^{-1}\left( \frac{- S_{\lleft}}{d_1} \right),\\
\xi_{\iin} &= \frac{S_{\rright}-S_{\lleft}}{d_1},\\
\xi_{\max} &= \xi_{\iin} + \sinh^{-1}\left( \frac{S_{\max} - S_{\rright}}{d_1} \right).
\end{align*}
Note that $\xi_{\min} < 0 < \xi_{\iin} < \xi_{\max}$.
The mesh $0=s_0 < s_1 < \ldots < s_{m_1}=S_{\max}$ is then defined
through the transformation
\begin{equation*}
s_i = \varphi(\xi_i) \quad (0\le i \le m_1)
\end{equation*}
where
\begin{equation*}
\varphi(\xi) =
\begin{cases}
 S_{\lleft} + d_1\sinh(\xi) & (\xi_{\min} \leq \xi < 0),\\
 S_{\lleft} + d_1\xi & (0 \leq \xi \leq \xi_{\iin} ),\\
 S_{\rright} + d_1\sinh(\xi-\xi_{\iin}) & (\xi_{\iin} < \xi \leq \xi_{\max}).
\end{cases}
\end{equation*}
\vskip0.2cm\noindent
This mesh for $s$ is uniform inside the interval $[S_{\lleft},S_{\rright}]$
and it is nonuniform outside.
The parameter $d_1$ controls the fraction of points $s_i$ that lie inside.
Put $\Delta \xi = \xi_1-\xi_0$.
It is readily seen that the above mesh is smooth, in the sense that there
exist real constants $C_0$, $C_1$, $C_2$ such that the mesh widths
$\Delta s_i = s_i -s_{i-1}$ satisfy
\begin{equation*}\label{smooth}
  C_0\, \Delta \xi \le \Delta s_i \le C_1\, \Delta \xi ~~ {\rm and} ~~
  |\Delta s_{i+1} - \Delta s_i| \le C_2 \left( \Delta \xi \right)^2 ~~
  ({\rm uniformly~in}~\, i,\, m_1).
\end{equation*}

\begin{figure}
\hskip1.5cm
\includegraphics[width=0.85\textwidth]{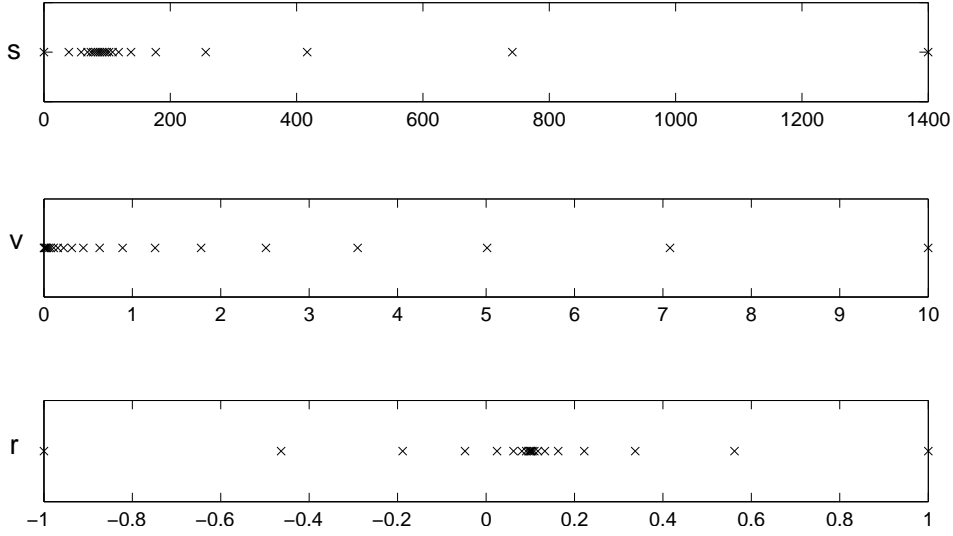}
\caption{Sample meshes for $s$, $v$, $r$ with $m_1=m_2=m_3=20$
and $K=100$, $T=1$, $c=0.1$.}\label{meshes}
\end{figure}

For the $v$- and $r$-directions we define nonuniform meshes of the
same type as considered in, eg, Tavella \& Randall (2000) and In 't Hout
\& Foulon (2010).
Let integers $m_2$, $m_3 \ge 1$ and parameters $c$, $d_2$, $d_3>0$ and
let equidistant points \mbox{$\eta_0 < \eta_1 < \ldots < \eta_{m_2}$}
and \mbox{$\zeta_0 < \zeta_1 < \ldots< \zeta_{m_3}$} be given by
\[
\eta_j = j\cdot \Delta \eta \quad (0\le j \le m_2)
\]
with
\[
\Delta \eta = \frac{1}{m_2} \sinh^{-1}(V_{\max}/d_2),
\]
and
\[
\zeta_k = \sinh^{-1}((-R_{\max}-c)/d_3) + k\cdot \Delta \zeta \quad
(0\le k \le m_3)
\]
with
\[
\Delta \zeta = \frac{1}{m_3}
\left[ \sinh^{-1}((R_{\max}-c)/d_3)-\sinh^{-1}((-R_{\max}-c)/d_3) \right].
\]
Then meshes $0=v_0 < v_1 < \ldots < v_{m_2}=V_{\max}$ and
$-R_{\max}=r_0 < r_1 < \ldots < r_{m_3}=R_{\max}$ are defined by
\[
v_j = d_2~{\rm sinh}(\eta_j) \quad (0\le j \le m_2)~~~{\rm and}~~~
r_k = c + d_3~{\rm sinh}(\zeta_k) \quad (0\le k \le m_3).
\]
It is easily verified that the meshes for $v$ and $r$ defined above are
also smooth.
The parameters $d_2$ and $d_3$ control, respectively, the fraction of
points $v_j$ that lie near $v=0$ and the fraction of points $r_k$ that
lie near a given interest rate level~$r=c$.
Here $c$ is chosen depending on the specific mean-reversion function~$b$.
For the $v$-mesh, besides the fact that the region $v\approx 0$ is of
practical importance, it is natural to place relatively many mesh points
there for numerical reasons, as the HHW PDE is convection-dominated in the
$v$-direction for $v\approx 0$ and the initial function is nonsmooth.

In this paper we set $S_{\max}=14K$, $V_{\max}=10$, $R_{\max}=1$.
This renders the error induced by the restriction of the spatial domain
of the HHW PDE to be negligible in our experiments.
Based on numerical tests, the parameters of the grid have been taken
equal to $d_1=K/20$, $d_2=V_{\max}/500$, $d_3=R_{\max}/400$ and, with
$r=\tfrac{1}{4}$,
\[
S_{\lleft}=\max\{\tfrac{1}{2},e^{-rT}\}K~,~~S_{\rright}=K.
\]
A further investigation into possibly better parameter values than
above may be interesting, but this is out of the scope of the
present paper.
Figure~\ref{meshes} displays sample meshes for the three spatial
directions if $m_1=m_2=m_3=20$ and $K=100$, $T=1$, $c=0.1$.
It is clear that the mesh points in the $s$-, $v$- and $r$-directions
are concentrated, respectively, near $s=K$, $v=0$ and $r=c$.

\subsection{Finite difference discretization}
Let $f:\R\rightarrow\R$ be any given function, let $\{ x_i \}_{i\in \Z}$
be any given increasing sequence of mesh points, and
$\Delta x_i = x_i-x_{i-1}$ for all $i$.
To approximate the first and second derivatives of $f$, we employ the
following well-known FD formulas:
\begin{subequations}
\begin{align}
f'(x_i) ~~\approx~~
&\alpha_{-2} f(x_{i-2}) +  \alpha_{-1}f(x_{i-1}) + \alpha_{0} f(x_{i}),\label{backw}\\
\nonumber\\
f'(x_i) ~~\approx~~
&\beta_{-1} f(x_{i-1}) + \beta_{0} f(x_i) + \beta_{1} f(x_{i+1}),\label{MethodB}\\
\nonumber\\
f'(x_i) ~~\approx~~
&\gamma_{0} f(x_{i}) + \gamma_{1} f(x_{i+1}) + \gamma_{2} f(x_{i+2}),\label{forw}\\
\nonumber\\
f''(x_i) ~~\approx~~
& \delta_{-1} f(x_{i-1}) + \delta_{0} f(x_i) + \delta_{1} f(x_{i+1})\label{diffusion}
\end{align}
\end{subequations}
with
\begin{align*}
\alpha_{-2}&= \tfrac{\Delta x_{i}}{\Delta x_{i-1} (\Delta x_{i-1} + \Delta x_{i})},
&\alpha_{-1}&= \tfrac{-\Delta x_{i-1}-\Delta x_i}{\Delta x_{i-1} \Delta x_{i}},
&\alpha_{0}&= \tfrac{\Delta x_{i-1} + 2\Delta x_i}{\Delta x_{i} (\Delta x_{i-1} + \Delta x_{i})},\\
\beta_{-1}&= \tfrac{-\Delta x_{i+1}}{\Delta x_i (\Delta x_i + \Delta x_{i+1})},
&\beta_{0}&= \tfrac{\Delta x_{i+1}-\Delta x_i}{\Delta x_i \Delta x_{i+1}},
&\beta_{1}&= \tfrac{\Delta x_i}{\Delta x_{i+1} (\Delta x_i + \Delta x_{i+1})},\\
\gamma_{0}&= \tfrac{-2\Delta x_{i+1}-\Delta x_{i+2}}{\Delta x_{i+1} (\Delta x_{i+1} + \Delta x_{i+2})},
&\gamma_{1}&= \tfrac{\Delta x_{i+1}+\Delta x_{i+2}}{\Delta x_{i+1} \Delta x_{i+2}},
&\gamma_{2}&= \tfrac{-\Delta x_{i+1}}{\Delta x_{i+2} (\Delta x_{i+1} + \Delta x_{i+2})},\\
\delta_{-1}&= \tfrac{2}{\Delta x_i (\Delta x_i + \Delta x_{i+1})},
&\delta_{0}&= \tfrac{-2}{\Delta x_i \Delta x_{i+1}},
&\delta_{1}&= \tfrac{2}{\Delta x_{i+1} (\Delta x_i + \Delta x_{i+1})}.
\end{align*}

Note that \eqref{MethodB} and \eqref{diffusion} are central schemes whereas
\eqref{backw} and \eqref{forw} are backward and forward schemes,
respectively.
If $f:\R^2\rightarrow \R$ is any given function of two variables $(x,y)$,
then we approximate the mixed derivative $f_{xy}(x_i,y_j)$ at any point
$(x_i,y_j)$ by successive application of the scheme \eqref{MethodB} in the
$x$- and $y$-directions.
This is equivalent to a FD formula based on a 9-point stencil centered
about $(x_i,y_j)$.
The FD schemes under consideration all possess a second-order truncation
error on smooth meshes whenever $f$ is sufficiently often continuously
differentiable.

The actual FD discretization of the initial-boundary value problem for
the HHW PDE is performed as follows.
In view of the Dirichlet conditions (\ref{BC}a) and (\ref{BC}c), the
relevant set of grid points is
\begin{align*}
\mathcal{G} = \{(s_i, v_j, r_k): 1\leq i \leq m_1\,,\, 0\leq j\leq m_2-1
\,,\, 0\leq k\leq m_3\}.
\end{align*}
At this grid, each spatial derivative appearing in (\ref{HHWPDE}) is
replaced by its corresponding central FD approximation, except:
\begin{itemize}
\item In the region $v>\eta$\, the backward scheme (\ref{backw}) is applied
for $\partial u /\partial v$.
This is done to alleviate spurious oscillations in the FD solution when
$\sigma_1$ is small.
It is well-known that such oscillations notably arise with central schemes 
if there is strong advection towards a Dirichlet boundary.
\item At the boundary $s=S_{\max}$ the derivatives in the $s$-direction
need to be considered.
The Neumann condition (\ref{BC}b) of course yields $\partial u /\partial s$
and it subsequently implies that $\partial^2 u/\partial s\partial v$ and
$\partial^2 u/\partial s \partial r$ vanish there.
Next, $\partial^2 u/ \partial s^2$ is approximated at
$s=s_{m_1}=S_{\max}$ by the scheme (\ref{diffusion}) with virtual point
$s_{m_1}+\Delta s_{m_1} > S_{\max}$ where the value at this point is
defined by linear extrapolation, using the value at $s_{m_1-1}$ and the
(given) derivative at $s_{m_1}$.
\item At the boundary $v=0$ we consider setting $v=0$ in the HHW PDE,
see Subsection \ref{BCs}.
Here $\partial u/\partial v$ is approximated using the forward scheme
(\ref{forw}).
We remark that this is done independently of whether or not the Feller
condition holds.
All other derivative terms in the $v$-direction vanish if $v=0$ and are
trivially dealt with.
\item At the boundaries $r = \pm R_{\max}$ the Neumann conditions
(\ref{BC}d) are incorporated similarly as for $s$ above.
\end{itemize}

The FD discretization of the initial-boundary value problem for the HHW PDE
leads to an initial value problem for a system of stiff ordinary differential
equations (ODEs),
\begin{align}\label{ODE}
U'(t) = A(t)U(t) + g(t) ~~~(0 \leq t \leq T),~~~U(0)=U_0.
\end{align}
Here $A(t)$, for $0\le t\le T$, is a given real square matrix and $g(t)$
is a given real vector that is determined by the boundary conditions.
The entries of the solution vector $U(t)$ form approximations to the
option values $u(s,v,r,t)$ at the spatial grid points
$(s,v,r) \in \mathcal{G}$, ordered in a convenient way.
The vector $U(0)=U_0$ is directly obtained by evaluation of the initial
function at $\mathcal{G}$.

We refer to (\ref{ODE}) as the {\it semidiscrete HHW PDE}.\,
The size of this system equals $M = m_1m_2(m_3+1)$ and is very large in
general.
In the experiments in this paper, we shall deal with sizes up to
approximately one million.


\setcounter{equation}{0}
\section{Time discretization: ADI schemes}\label{ADI}
Selecting a suitable time discretization scheme for the semidiscrete
HHW PDE (\ref{ODE}) is the key to obtaining an effective full numerical
solution method for the HHW initial-boundary value problem.
Popular standard methods such as the Crank--Nicolson scheme are often not
efficient anymore.
The reason for this lies in the fact that in each new time step very large
systems of linear equations need to be solved involving the matrix $A(t)$
for one or more new values of~$t$.
Due to its large bandwidth, this is computationally very demanding.

For the time discretization of the semidiscrete HHW PDE, we consider in
the present paper splitting schemes of the ADI type.
Here, the matrix $A(t)$ is decomposed into four simpler matrices,
\begin{equation*}\label{split}
A(t) = A_0 + A_1 + A_2 + A_3(t).
\end{equation*}
The matrix $A_0$ represents the part of $A(t)$ that stems from the FD
discretization of all mixed derivative terms in the HHW PDE.
Note that $A_0$ is nonzero whenever at least one of the correlation
factors $\rho_{12}$, $\rho_{13}$, $\rho_{23}$ is nonzero.
In line with the classical ADI idea, the matrices $A_1$, $A_2$, $A_3(t)$
represent the parts of $A(t)$ that stem from the FD discretization of all
spatial derivatives in the $s$-, $v$- and $r$-directions, respectively.
The $ru$ term in (\ref{HHWPDE}) is distributed evenly over $A_1$, $A_2$,
$A_3(t)$.
We decompose $g(t) = g_0 + g_1 + g_2 + g_3(t)$ analogously to $A(t)$.
The matrices $A_1$, $A_2$, $A_3(t)$ are essentially tridiagonal,
pentadiagonal and tridiagonal, respectively.
Note that the time-dependency of $A(t)$ is only passed on to the matrix
$A_3(t)$, ie, the matrices $A_0$, $A_1$, $A_2$ are time-independent.

Let $\theta >0$ be a given real parameter and $\Delta t = T/N$ with
integer $N\ge 1$.
Set $t_n = n\,\Delta t$ and
$\Delta g_n = g_3\left(t_n\right) - g_3\left(t_{n-1}\right)=
g\left(t_n\right) - g\left(t_{n-1}\right)$.
We study four ADI schemes which all generate, in a one-step manner,
successive approximations $U_n$ to the solution vectors $U(t_n)$ of
(\ref{ODE}) for $n=1,2,\ldots,N$.
\vskip0.5cm\noindent
\textit{Douglas (Do) scheme:}
\begin{equation}\label{Do}
\left\{\begin{array}{l}
Y_0 = U_{n-1}+\Delta t\, \big( A\left(t_{n-1}\right)U_{n-1}+ g\left(t_{n-1}\right)\big),
\vspace*{0.2cm}\\
Y_j = Y_{j-1}+\theta\,\Delta t\, A_j \big(Y_j-U_{n-1}\big)~~~(j=1,2),
\vspace*{0.2cm}\\
Y_3 = Y_{2}+\theta\,\Delta t\, \big(A_3\left(t_n\right)Y_3 -A_3\left(t_{n-1}\right)U_{n-1}
+\Delta g_n \big),\phantom{xxx}
\vspace*{0.2cm}\\
U_n = Y_3.
\end{array}\right.
\end{equation}
\vskip0.3cm\noindent
\textit{Craig--Sneyd (CS) scheme:}
\begin{equation}\label{CS}
\left\{\begin{array}{l}
Y_0 = U_{n-1}+\Delta t\, \big( A\left(t_{n-1}\right)U_{n-1}+ g\left(t_{n-1}\right)\big),
\vspace*{0.2cm}\\
Y_j = Y_{j-1}+\theta\, \Delta t\, A_j \big(Y_j-U_{n-1}\big)~~~(j=1,2),
\vspace*{0.2cm}\\
Y_3 = Y_{2}+\theta\,\Delta t\, \big(A_3\left(t_n\right)Y_3 -A_3\left(t_{n-1}\right)U_{n-1}+
\Delta g_n \big),\phantom{xxx}
\vspace*{0.2cm}\\
\widetilde{Y}_0 = Y_0+\tfrac{1}{2}\Delta t\, A_0 \big(Y_3-U_{n-1}\big),
\vspace*{0.2cm}\\
\widetilde{Y}_j = \widetilde{Y}_{j-1}+\theta\,\Delta t\, A_j \big(\widetilde{Y}_j-U_{n-1}\big)~~~(j=1,2),
\vspace*{0.2cm}\\
\widetilde{Y}_3 = \widetilde{Y}_{2}+\theta\,\Delta t\, \big(A_3\left(t_n\right)\widetilde{Y}_3
-A_3\left(t_{n-1}\right)U_{n-1}+\Delta g_n \big),
\vspace*{0.2cm}\\
U_n = \widetilde{Y}_3.
\end{array}\right.
\end{equation}
\vskip0.3cm\noindent
\textit{Modified Craig--Sneyd (MCS) scheme:}
\begin{equation}\label{MCS}
\left\{\begin{array}{l}
Y_0 = U_{n-1}+\Delta t\, \big( A\left(t_{n-1}\right)U_{n-1}+ g\left(t_{n-1}\right)\big),
\vspace*{0.2cm}\\
Y_j = Y_{j-1}+\theta\,\Delta t\, A_j \big(Y_j-U_{n-1}\big)~~~(j=1,2),
\vspace*{0.2cm}\\
Y_3 = Y_{2}+\theta\,\Delta t\, \big(A_3\left(t_n\right)Y_3 -A_3\left(t_{n-1}\right)U_{n-1}+
\Delta g_n \big),
\vspace*{0.2cm}\\
\widehat{Y}_0 = Y_0+\theta\, \Delta t\, A_0 \big(Y_3-U_{n-1}\big),
\vspace*{0.2cm}\\
\widetilde{Y}_0 = \widehat{Y}_0+\left(\tfrac{1}{2}-\theta\right) \Delta t\,
\big(A\left(t_n\right)Y_3 -A\left(t_{n-1}\right)U_{n-1}+ \Delta g_n\big),
\vspace*{0.2cm}\\
\widetilde{Y}_j = \widetilde{Y}_{j-1}+\theta\,\Delta t\, A_j
\big(\widetilde{Y}_j-U_{n-1}\big)~~~(j=1,2),
\vspace*{0.2cm}\\
\widetilde{Y}_3 = \widetilde{Y}_{2}+\theta\,\Delta t\, \big(A_3\left(t_n\right)\widetilde{Y}_3
-A_3\left(t_{n-1}\right)U_{n-1}+\Delta g_n \big),
\vspace*{0.2cm}\\
U_n = \widetilde{Y}_3.
\end{array}\right.
\end{equation}
\vskip0.3cm\noindent
\textit{Hundsdorfer--Verwer (HV) scheme:}
\begin{equation}\label{HV}
\left\{\begin{array}{l}
Y_0 = U_{n-1}+\Delta t\, \big( A\left(t_{n-1}\right)U_{n-1}+ g\left(t_{n-1}\right)\big),
\vspace*{0.2cm}\\
Y_j = Y_{j-1}+\theta\,\Delta t\, A_j \big(Y_j-U_{n-1}\big)~~~(j=1,2),
\vspace*{0.2cm}\\
Y_3 = Y_{2}+\theta\,\Delta t\, \big(A_3\left(t_n\right)Y_3 -A_3\left(t_{n-1}\right)U_{n-1}
+\Delta g_n \big),
\vspace*{0.2cm}\\
\widetilde{Y}_0 = Y_0+\tfrac{1}{2}\Delta t\, \big(A\left(t_n\right)Y_3 -A\left(t_{n-1}\right)
U_{n-1}+ \Delta g_n\big),\phantom{xxxxx}
\vspace*{0.2cm}\\
\widetilde{Y}_j = \widetilde{Y}_{j-1}+\theta\,\Delta t\, A_j \big(\widetilde{Y}_j-Y_3\big)~~~(j=1,2),
\vspace*{0.2cm}\\
\widetilde{Y}_3 = \widetilde{Y}_{2}+\theta\,\Delta t\, A_3\left(t_n\right)
\big(\widetilde{Y}_3  -Y_{3} \big),\vspace*{0.2cm}\\
U_n = \widetilde{Y}_3.
\end{array}\right.
\end{equation}
\vskip0.2cm
The CS, MCS, HV schemes can be viewed as different extensions to the Do scheme.
The CS and MCS schemes are equivalent if (and only if) $\theta=\tfrac{1}{2}$.

It is readily observed that in the four ADI schemes the $A_0$ part,
representing all mixed derivatives in the HHW PDE, is always treated
in an {\it explicit}\, fashion.
The first papers to propose this kind of adaptation of the classical ADI
schemes to PDEs with mixed derivative terms are, to our knowledge, McKee
\& Mitchell (1970) and Craig \& Sneyd (1988).

Following the classical ADI approach, the $A_1$, $A_2$, $A_3(t)$ parts
are treated in an {\it implicit}\, fashion.
In every step of each scheme, systems of linear equations need to be
solved, successively involving the matrices $(I-\theta\, \Delta t\, A_j)$
for $j=1, 2$ and $(I-\theta\, \Delta t\, A_3(t_n))$, where $I$ is the
identity matrix.
As all these matrices have a fixed, small bandwidth (of at most five)
this can be done efficiently by $LU$ factorization.
Note that for $j=1, 2$ the pertinent matrices are further independent
of the step index $n$, and hence, their $LU$ factorizations can be
computed once, beforehand, and then used in all time steps.

By Taylor expansion one obtains (after some elaborate calculations) the
classical order of consistency of each ADI scheme, ie, the order of
consistency in the nonstiff sense.
For any given~$\theta$, the order of the Do scheme is just one if $A_0$
is nonzero.
This low order is due to the fact that the $A_0$ part is treated in a
simple, explicit Euler fashion.
The CS scheme has order two provided $\theta=\tfrac{1}{2}$.
The MCS and HV schemes are of order two for any given $\theta$.
With the latter schemes, the parameter $\theta$ can thus be chosen
to meet additional requirements.

A detailed discussion, with ample references to the literature, concerning
the above four ADI schemes has been given in In 't Hout \& Welfert (2007,
2009).
The Do and CS schemes are already often applied to PDEs in finance, see eg,
Andersen \& Andreasen (2000), Lipton (2001), Randall (2002) and Andersen
\& Piterbarg (2010).
More recently, the MCS and HV schemes have gained interest, see eg,
In 't Hout (2007), Dang, Christara, Jackson \& Lakhany (2010), In 't Hout
\& Foulon (2010), Haentjens \& In 't Hout (2010), Egloff (2011) and Itkin
\& Carr (2011).

For an effective application of numerical schemes, stability is imperative.
The stability of ADI schemes in the case of PDEs possessing mixed derivative
terms has been analyzed by a number of authors in the literature.
This stability analysis has been performed in the von Neumann (Fourier)
framework.
Here one considers application to the semidiscretized convection-diffusion
equation
\begin{equation*}\label{cd}
\frac{\partial u}{\partial t}=
\mathbf{c}\cdot\nabla u + \nabla \cdot (D\nabla u)
\end{equation*}
on a rectangular domain, with constant real vector $\mathbf{c}$ and constant,
positive semidefinite real matrix $D=(d_{ij})$, with periodic boundary
condition, on a uniform spatial grid, and one studies stability in the
$l_2$-norm.
Note that the presence of mixed derivative terms corresponds to the matrix
$D$ being nondiagonal.
A desirable property is {\it unconditional}\, stability, ie, without any
restriction on the time step $\Delta t >0$.

The most comprehensive stability results for the Do, CS, MCS and HV schemes
in the literature up to now, relevant to PDEs with mixed derivative terms,
are given in In 't Hout \& Welfert (2007, 2009), In 't Hout \& Mishra (2010,
2011).
We review the main conclusions from loc cit pertinent to two and three
spatial dimensions.
Here stability is always understood in the von Neumann sense and
unconditional.
To formulate some of the results, we consider for $\gamma \in [0,1]$
the following condition on $D$,
\begin{equation}\label{gamma}
|d_{ij}|\le \gamma\, \sqrt{d_{ii}d_{jj}} \quad \text{for~all}~~i\not= j.
\end{equation}
The quantity $\gamma$ can be viewed as a measure for the relative size of
the mixed derivative coefficients.
Because $D$ is positive semidefinite, the condition (\ref{gamma}) is
always fulfilled with $\gamma =1$.
But in actual applications, in particular the HHW PDE, one usually has
more information, namely $\gamma<1$.

For two-dimensional convection-diffusion equations with mixed derivative
term, the Do and CS schemes are both stable whenever $\theta\ge\tfrac{1}{2}$.
If there is no convection ($\mathbf{c}=\mathbf{0}$), then the MCS
and HV schemes are stable whenever $\theta\ge\tfrac{1}{3}$ and
$\theta\ge 1-\tfrac{1}{2}\sqrt{2}~\,(\approx 0.293)$, respectively.
For the MCS scheme, stability has been proved for general two-dimensional
equations, with convection, if \mbox{$\tfrac{1}{2}\le\theta\le 1$.}
Next, based on strong numerical evidence, stability of the MCS scheme
for the special value $\theta=\tfrac{1}{3}$ was conjectured under the
mild, additional condition that (\ref{gamma}) holds with $\gamma \le 0.96$.
For the HV scheme, stability for general two-dimensional equations has
been conjectured for all
$\theta\ge\tfrac{1}{2}+\tfrac{1}{6}\sqrt{3}~\,(\approx 0.789)$.
We note that the latter bound stems from Lanser, Blom \& Verwer (2001),
who proved it to be necessary and sufficient for stability in the case
of two-dimensional equations without mixed derivatives.

For three-dimensional problems, positive results on the stability of the
ADI schemes have been derived for pure diffusion equations with mixed
derivative terms.
In this case, it has been shown that the Do, CS, MCS and HV schemes are
stable whenever
\mbox{$\theta\ge\tfrac{2}{3}$},\, $\theta\ge\tfrac{1}{2}$,\,
$\theta\ge\max\{\tfrac{1}{4},\tfrac{2}{13}(2\gamma+1)\}$ and
$\theta\ge\tfrac{3}{2}(2-\sqrt{3})~\,(\approx 0.402)$,
respectively.\footnote{The result for the Do scheme is new;
its proof will be included in a forthcoming paper.}

At this moment sufficient conditions on $\theta$ for stability of
the ADI schemes pertinent to general three-dimensional
convection-diffusion equations with mixed derivative terms are
lacking in the literature. Accordingly, we select the parameters
$\theta$, in the subsequent experiments, on the basis of the present
results, reviewed above.

In practical applications it turns out that a smaller value $\theta$
often leads to a smaller error constant. In view of this, we choose
$\theta$ as small as possible under the requirement of
(unconditional) stability.


\setcounter{equation}{0}
\section{Numerical experiments}\label{numexp}
In this section we present extensive numerical tests with the four ADI
schemes \eqref{Do}, \eqref{CS}, \eqref{MCS}, \eqref{HV} in the application
to the semidiscrete HHW PDE described in Section \ref{space}.
This yields important insight in their actual stability and convergence
behavior and mutual performance.
We consider the HHW model with arbitrary (nonzero) correlation factors,
with time-dependent mean-reversion levels, for cases where the Feller
condition is satisfied and for cases where it is not.
In addition, we deal with European call options as well as up-and-out
call options.

For the diffusion matrix of the HHW PDE,
\[
D(s,v) =
\frac{1}{2}
\left(
  \begin{array}{ccc}
    s^2v                          & \rho_{12} \sigma_1 s v               & \rho_{13} \sigma_2 s \sqrt{v} \\
    \rho_{12} \sigma_1 s v        & \sigma_1^2 v                         & \rho_{23} \sigma_1 \sigma_2 \sqrt{v} \\
    \rho_{13} \sigma_2 s \sqrt{v} & \rho_{23} \sigma_1 \sigma_2 \sqrt{v} & \sigma_2^2 \\
  \end{array}
\right),
\]
it is easily verified that the condition (\ref{gamma}) holds with
$\gamma =\max \left\{ |\rho_{12}|, |\rho_{13}|, |\rho_{23}| \right\}$.
Based on the stability and accuracy results discussed in Section~\ref{ADI}
we select, for this value~$\gamma$,
\begin{itemize}
  \item the Do~~~\,\,scheme \eqref{Do}\, with $\theta = \frac{2}{3}$
  \item the CS~~~\,\,scheme \eqref{CS}\, with $\theta = \frac{1}{2}$
  \item the MCS\, scheme \eqref{MCS}\,   with $\theta = \max\{\frac{1}{3},\frac{2}{13}(2\gamma+1)\}$
  \item the HV~~~\,scheme \eqref{HV}\,   with $\theta = \frac{1}{2}+\frac{1}{6}\sqrt{3}$.
\end{itemize}
The Do scheme has classical order one and the CS, MCS, HV schemes all
possess classical order two.
Note that the MCS scheme has $\theta\le \tfrac{6}{13}~\,(\approx 0.462)$.

For the ADI schemes under consideration we shall study in this section
the \textit{global temporal discretization error}, defined by
\begin{equation}\label{et}
\widehat{e}\,(\Delta t;m_1,m_2,m_3) =
\max\{\,|U_l(T)-U_{N,l}|:\,(s_i, v_j, r_k) \in {\cal D}\,\},
\end{equation}
where $T=N\Delta t$ with integer $N\ge 1$ and $U(T)$ denotes the
exact solution vector to the semidiscrete HHW PDE (\ref{ODE})
at time $T$.
The index $l = l(i,j,k)$ corresponds to the spatial grid point
$(s_i, v_j, r_k)$ and ${\cal D}$ is a natural region of interest,
to be specified below.

If $\rho_{13}=\rho_{23}=0$, then a semi closed-form analytic formula
for European call option values is known, see the Appendix.
We shall employ this formula to validate the FD discretization of the
HHW PDE from Section~\ref{space} and to study the
\textit{global spatial discretization error} in this case, defined by
\begin{equation}\label{es}
e(m_1,m_2,m_3) =
\max\{\,|u(s_i,v_j,r_k,T)-U_l(T)|:\,(s_i, v_j, r_k) \in {\cal D}\,\}.
\end{equation}

The temporal and spatial discretization errors are both measured in the
maximum norm, which is highly relevant to financial applications.
In order to compute (\ref{et}) and (\ref{es}) for a given spatial grid,
we use a sufficiently accurate reference value for $U(T)$, obtained by 
applying the MCS scheme to (\ref{ODE}) with $N=20000$ and $N=200$ time 
steps, respectively.

For efficiency of the spatial discretization it turns out that one can
place relatively less grid points in the $v$- and $r$-directions than
in the $s$-direction.
Accordingly, we choose in the following the numbers of grid points in
the three spatial directions as $m_1=2m$, $m_2=m_3=m$ with integer~$m$.
Note that the size of the semidiscrete HHW system equals $M=2m^2(m+1)$.

We are interested in mean-reversion levels $b$ that are time-dependent.
As an example, we choose
\begin{equation}\label{meanr}
b(\tau)=c_1-c_2e^{-c_3\tau} \quad (\tau \ge 0)
\end{equation}
with positive constants $c_1$, $c_2$, $c_3$ and $c_1>c_2$.
This choice for~$b$ is somewhat arbitrary, but the conclusions obtained
below on the numerical schemes are the same for other (more realistic)
time-dependent mean-reversion levels.
For the mesh in the $r$-direction, defined in Subsection \ref{grid},
we take $c=c_1$.

\begin{table}
\begin{center}
\begin{tabular}{|c|l|l|l|l|l|l|}
        \hline
        & Case A & Case B & Case C & Case D & Case E & Case F\\
        \hline \hline
        $\kappa$    & 3    & 0.6067 & 2.5  & 0.5   & 0.3  & 1\\
        $\eta$      & 0.12 & 0.0707 & 0.06 & 0.04  & 0.04 & 0.09\\
        $\sigma_1$  & 0.04 & 0.2928 & 0.5  & 1     & 0.9  & 1\\
        \hline
        $a$         & 0.2  & 0.05   & 0.15  & 0.08  & 0.16  & 0.22\\
        $c_1$       & 0.05 & 0.055  & 0.101 & 0.103 & 0.055 & 0.074\\
        $c_2$       & 0.01 & 0.005  & 0.001 & 0.003 & 0.025 & 0.014\\
        $c_3$       & 1    & 4      & 2.3   & 1     & 1.6   & 2.1\\
        $\sigma_2$  & 0.03 & 0.06   & 0.1   & 0.09  & 0.03  & 0.07\\
        \hline
        $\rho_{12}$ & 0.6     & -0.7571             & -0.1
    & -0.9               & -0.5               & -0.3    \\
        $\rho_{13}$ & 0.2 (0) & \phantom{-}0.6 (0)  & -0.3 (0)
    & \phantom{-}0.6 (0) & \phantom{-}0.2 (0) & -0.5 (0)\\
        $\rho_{23}$ & 0.4 (0) & -0.2 (0)            & \phantom{-}0.2 (0)
    & -0.7 (0)           & \phantom{-}0.1 (0) & -0.2 (0)\\
        \hline
        $T$ & 1   & 3   & 0.25 & 10  & 15  & 5\\
        $K$ & 100 & 100 & 100  & 100 & 100 & 100\\
        \hline
\end{tabular}
\end{center}
\caption{Parameters for the Heston--Hull--White model.}
\label{cases1}
\end{table}

\subsection{European call options}\label{Eurcall}
Our first experiments concern European call option values in the six
cases of parameter sets for the HHW model listed in Table~\ref{cases1}.

The cases A, B, C can be viewed as an extension of three test cases
for the Heston model previously used in In 't Hout \& Foulon (2010).
The values $\kappa$, $\eta$, $\sigma_1$, $\rho_{12}$, $T$ stem from
Bloomberg (2005), Schoutens, Simons \& Tistaert (2004) and Winkler,
Apel \& Wystup (2002), respectively.
Here the Feller condition always holds.

The cases D, E, F form an extension of the three cases for the Heston
model presented by Andersen (2008).
They are proposed in loc cit as challenging test cases for practical
applications.
Notably, the Feller condition is {\it not}\, fulfilled.
Also, the maturity times are large.

In all six cases, the values $a$, $c_1$, $c_2$, $c_3$, $\sigma_2$
pertinent to the Hull--White model as well as the two correlations
$\rho_{13}$, $\rho_{23}$ are chosen in an arbitrary, realistic way.
Here the corresponding correlation matrices are always positive
definite.

\begin{figure}
\begin{center}
\begin{tabular}{c c}
         \includegraphics[width=0.5\textwidth]{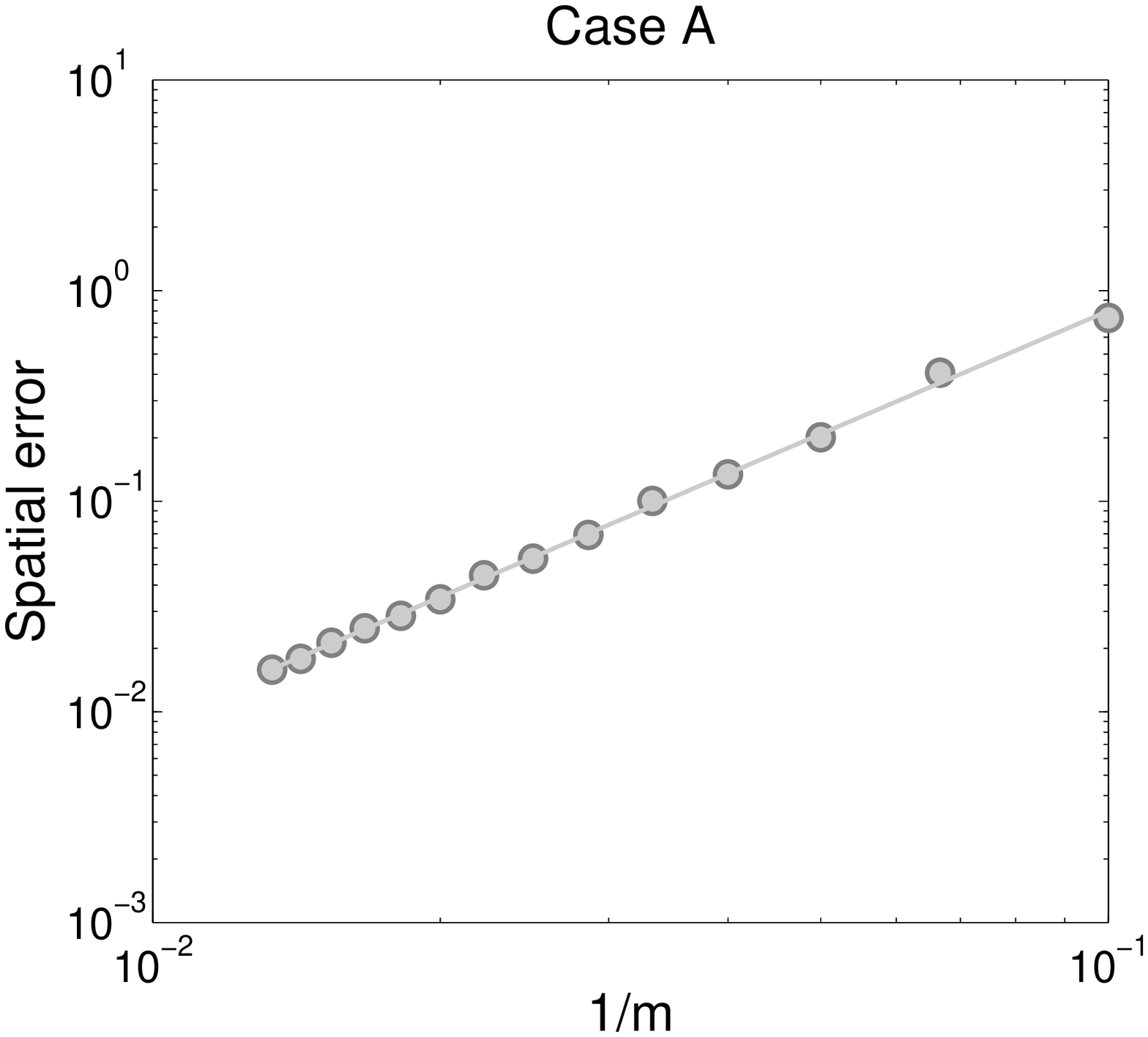}&
         \includegraphics[width=0.5\textwidth]{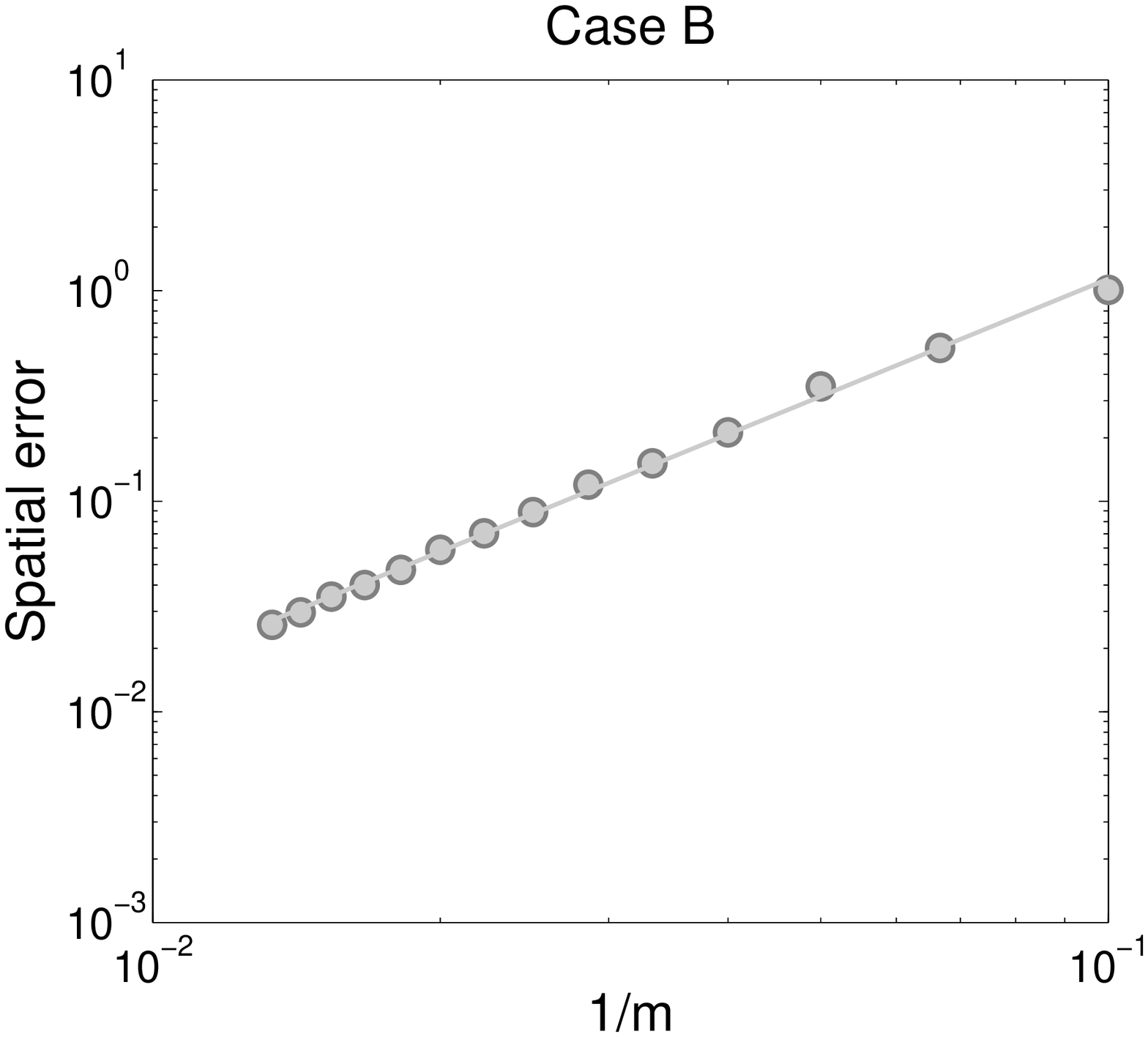}\\
         \includegraphics[width=0.5\textwidth]{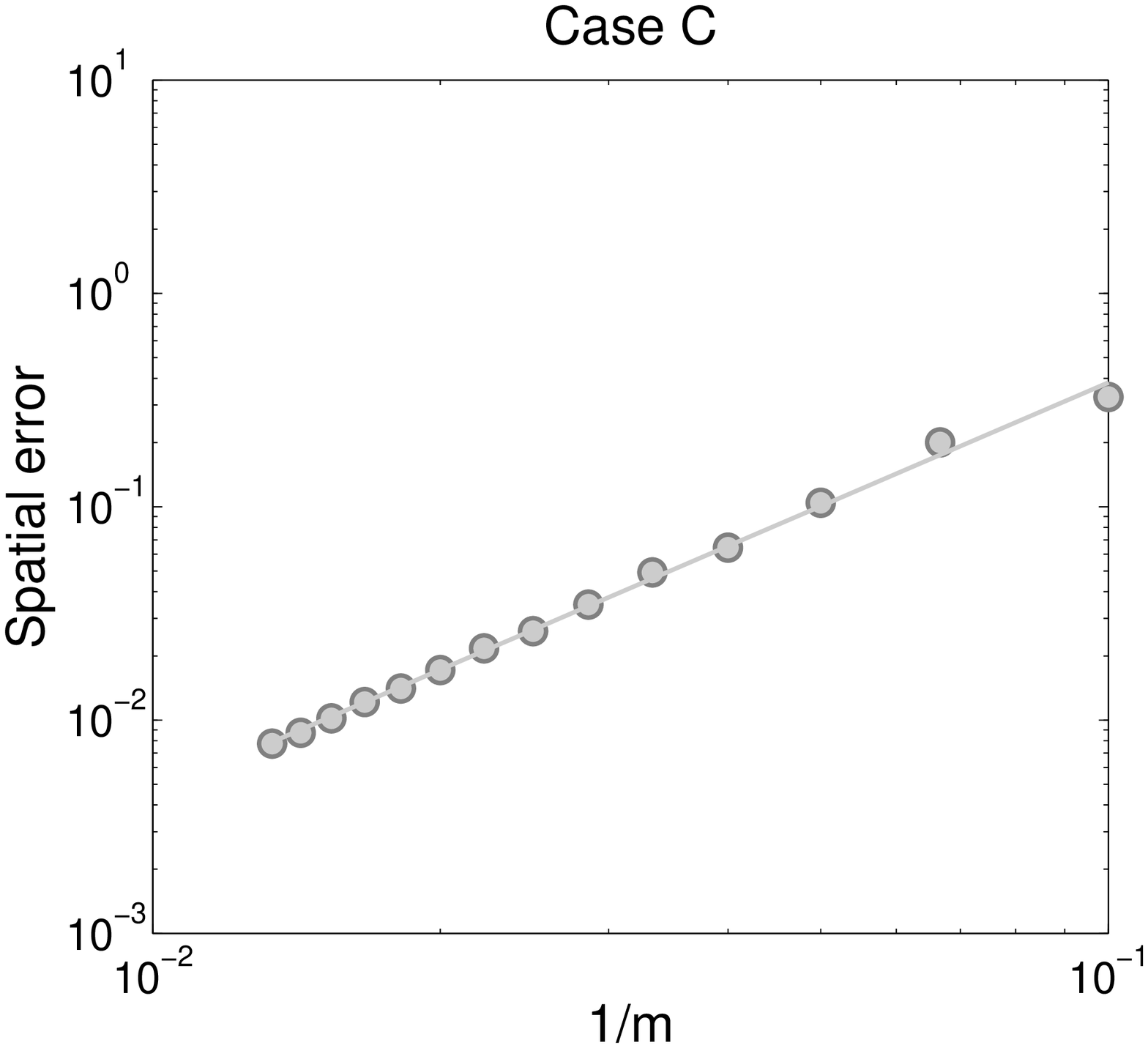}&
         \includegraphics[width=0.5\textwidth]{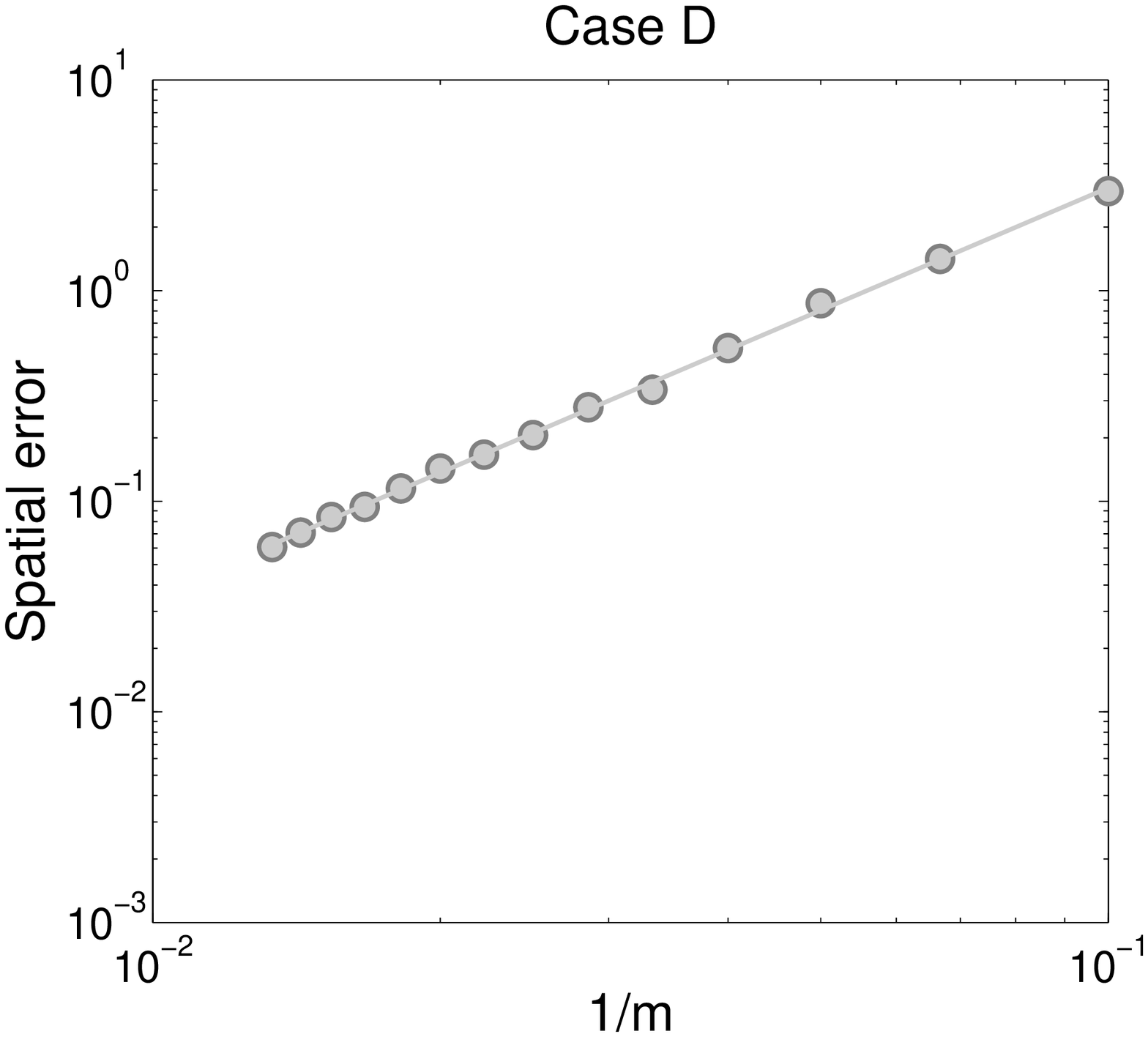}\\
         \includegraphics[width=0.5\textwidth]{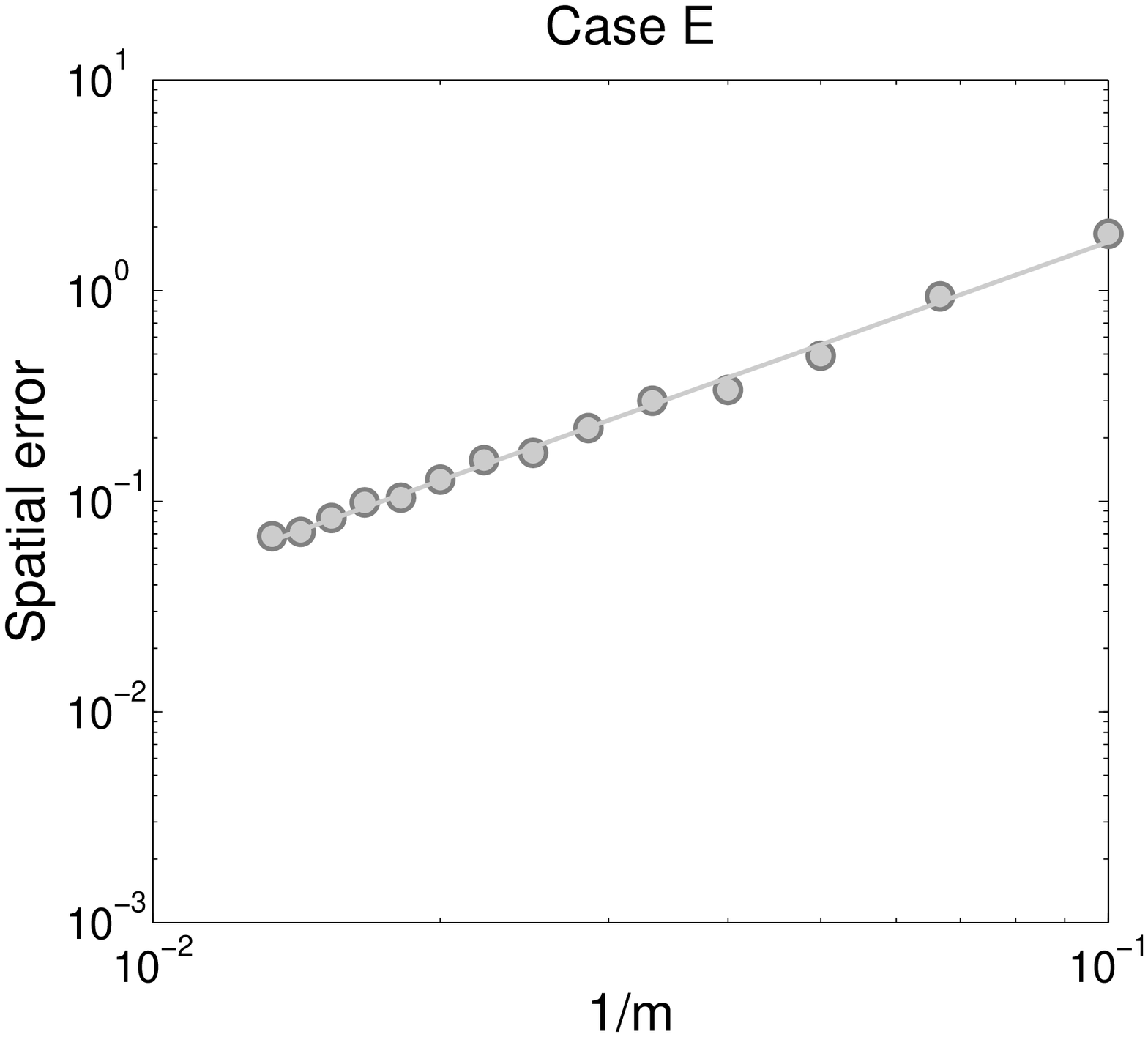}&
         \includegraphics[width=0.5\textwidth]{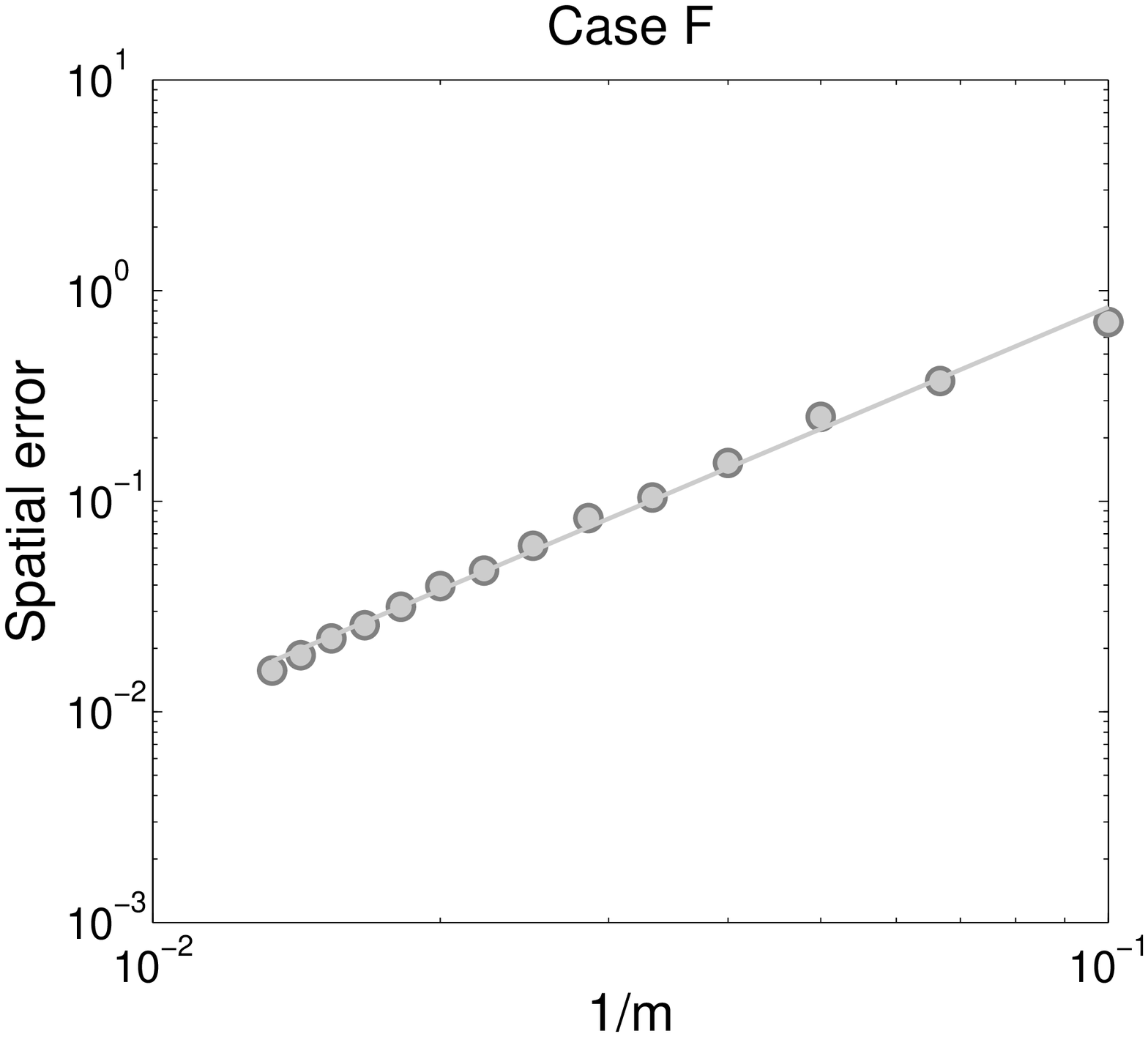}
\end{tabular}
\end{center}
\caption{Spatial discretization errors $e(2m,m,m)$ vs $1/m$ for
European call options in the six cases of Table~\ref{cases1} with
$\rho_{13}=\rho_{23}=0$ for $m=10,15,\ldots,75$.
\vspace{9mm}
}
\label{SpatialError1}
\end{figure}

\begin{figure}
\begin{center}
\begin{tabular}{c c}
         \includegraphics[width=0.5\textwidth]{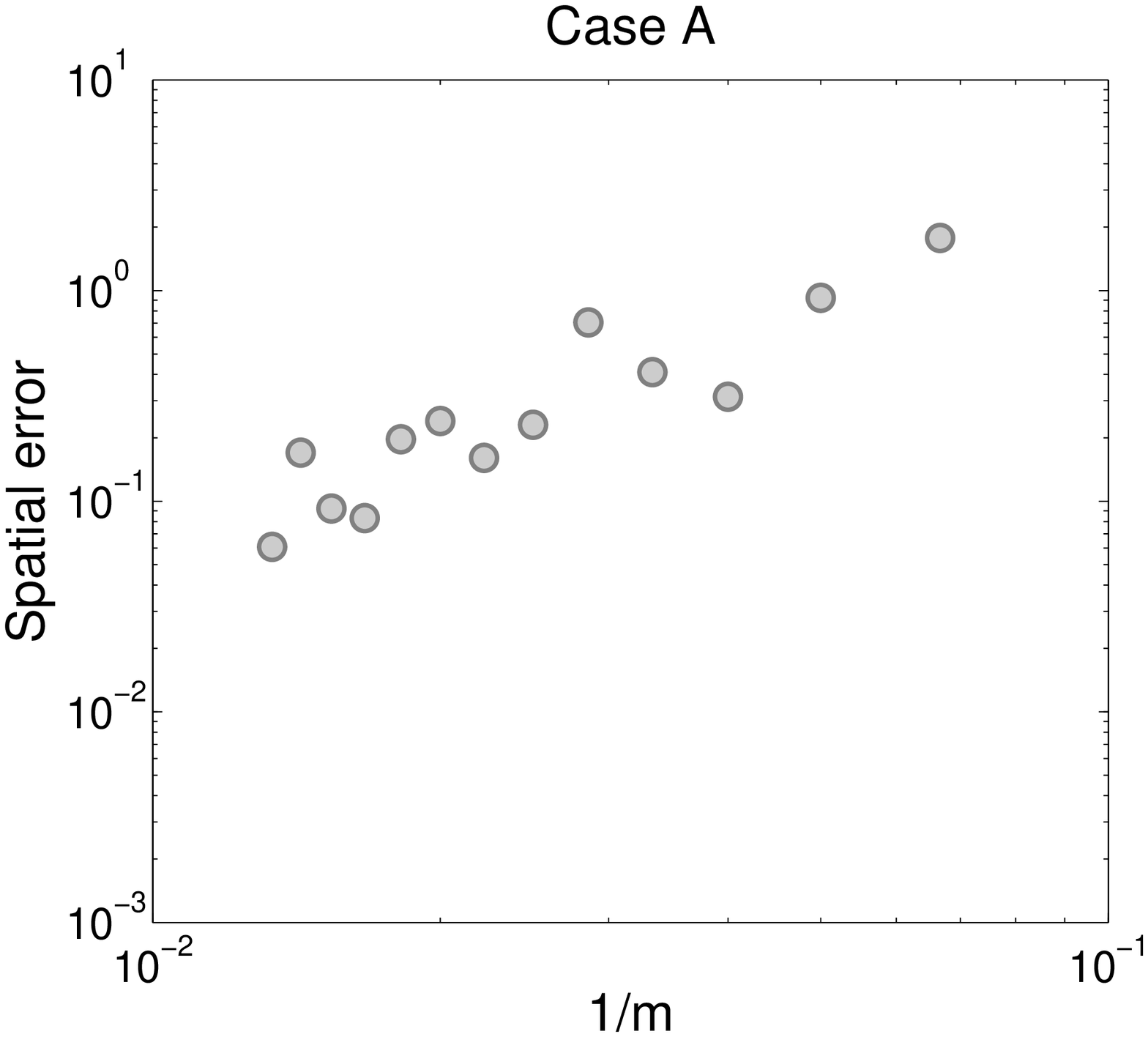}&
         \includegraphics[width=0.5\textwidth]{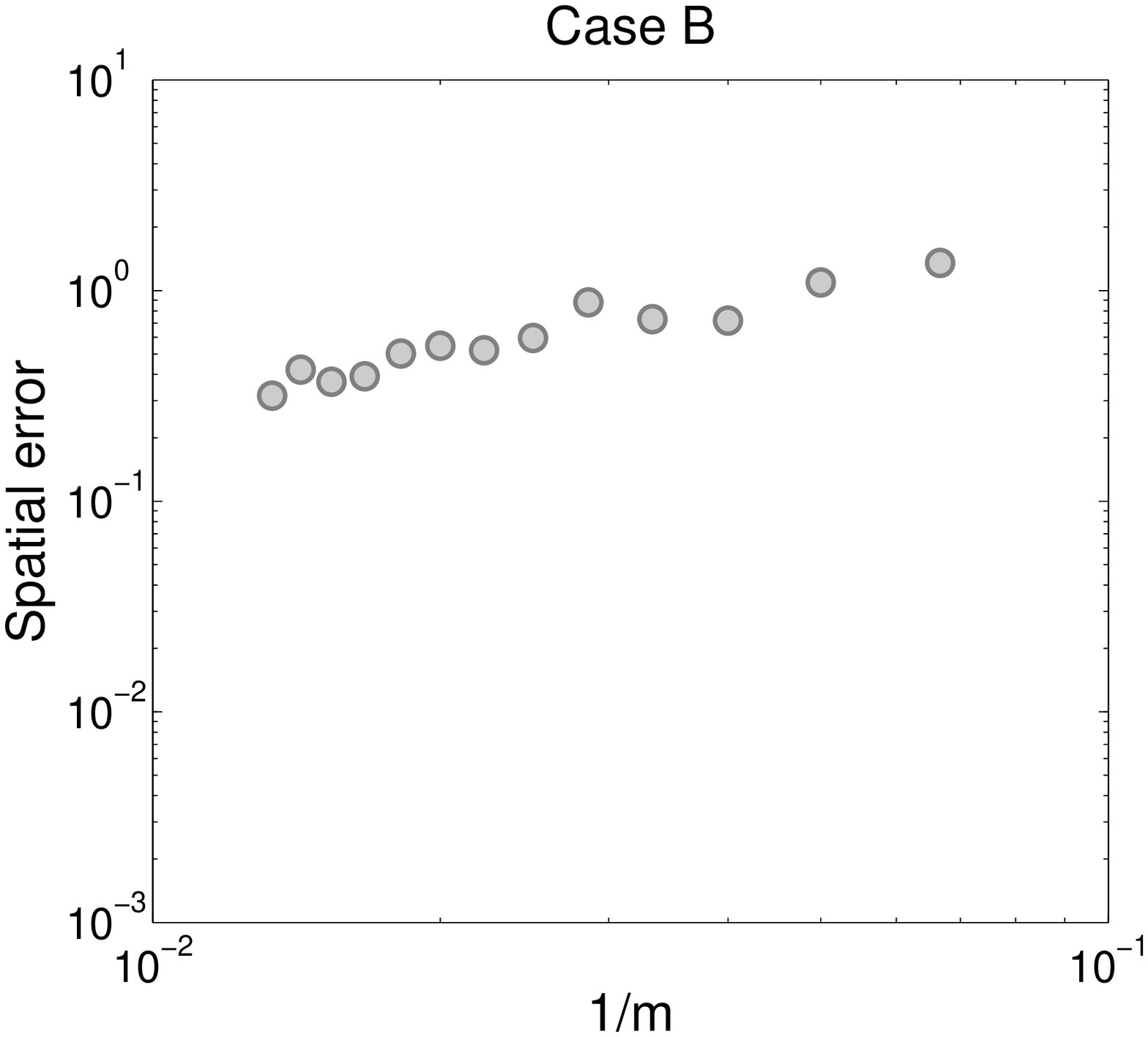}\\
         \includegraphics[width=0.5\textwidth]{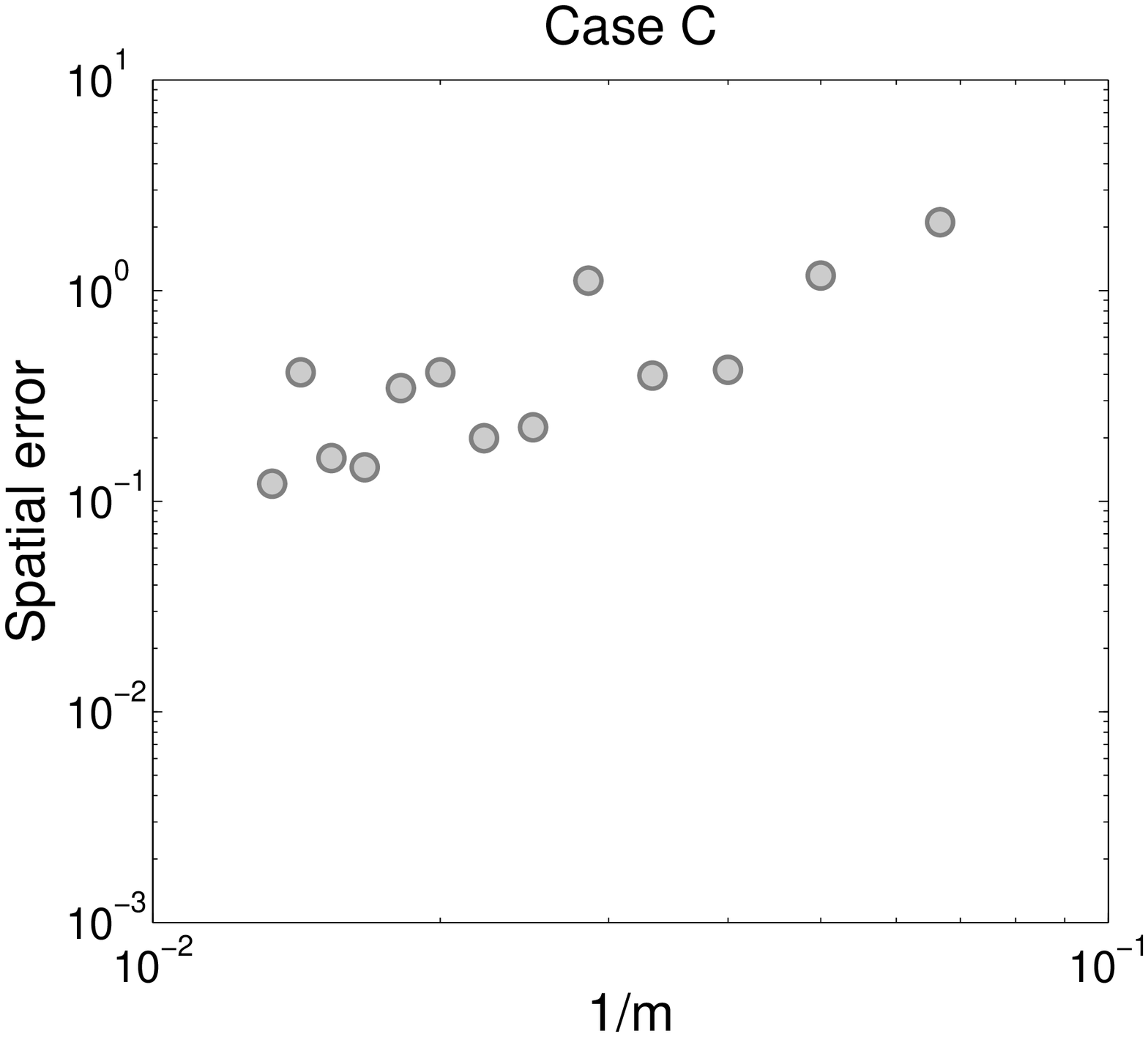}&
         \includegraphics[width=0.5\textwidth]{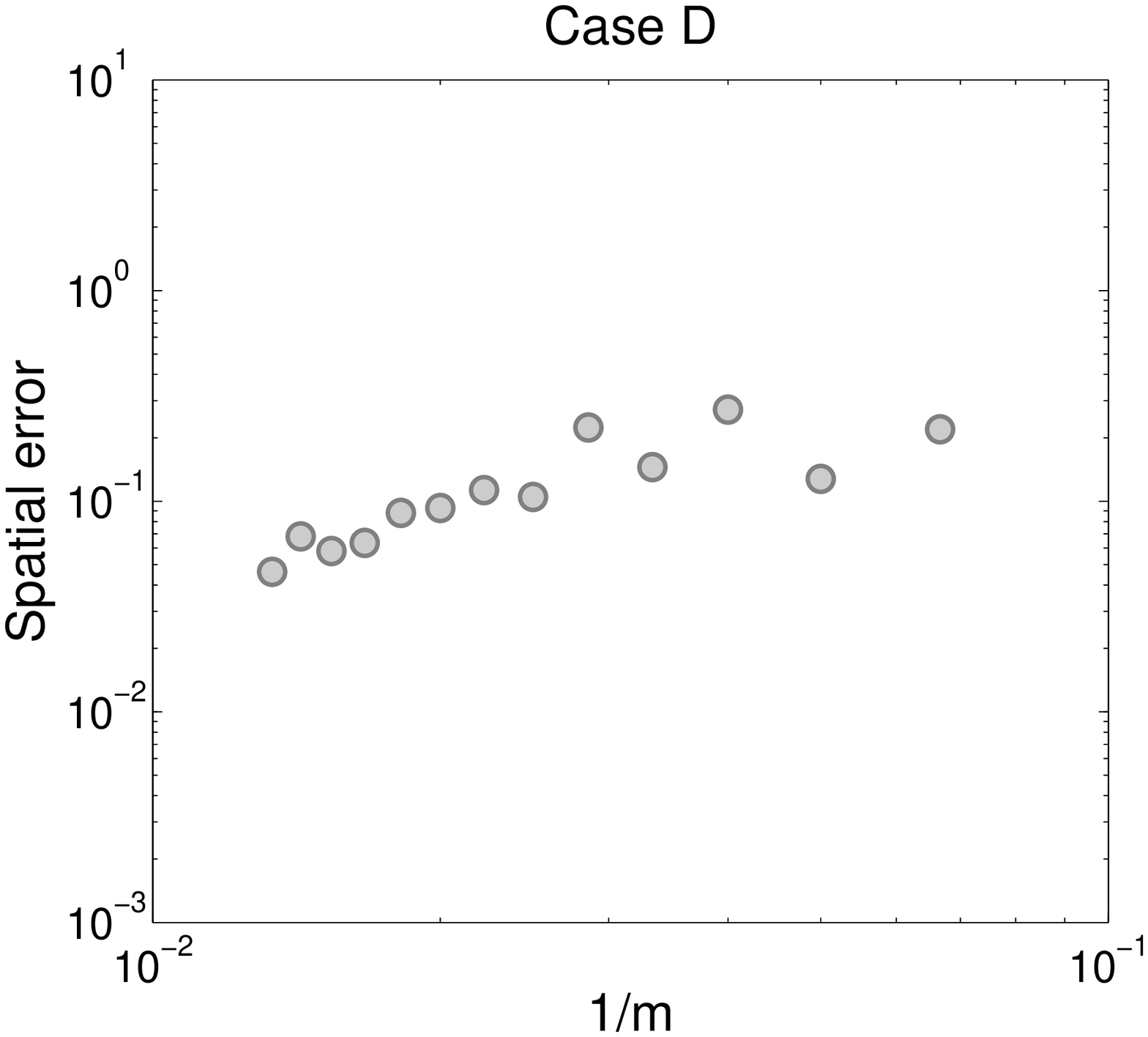}\\
         \includegraphics[width=0.5\textwidth]{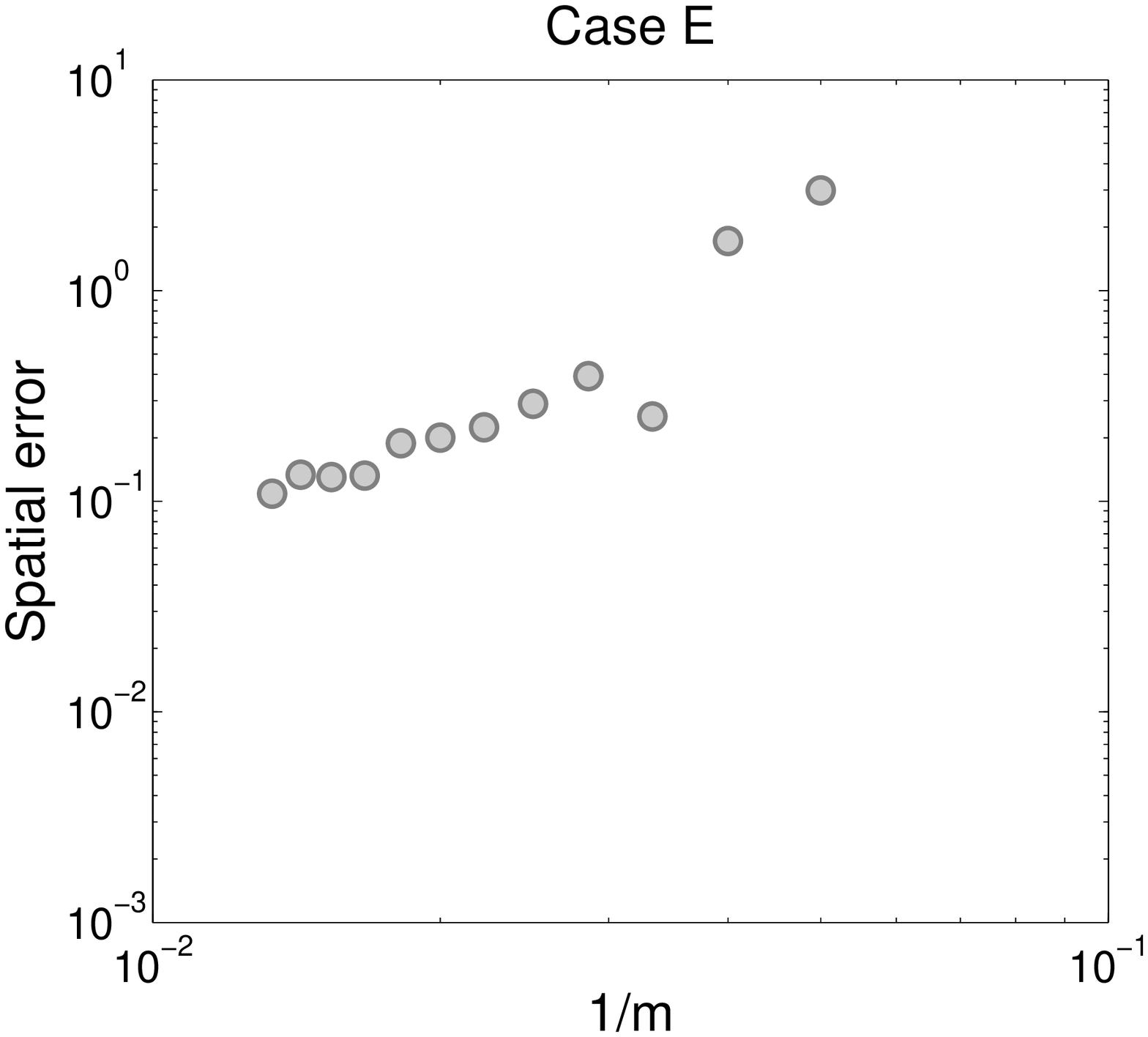}&
         \includegraphics[width=0.5\textwidth]{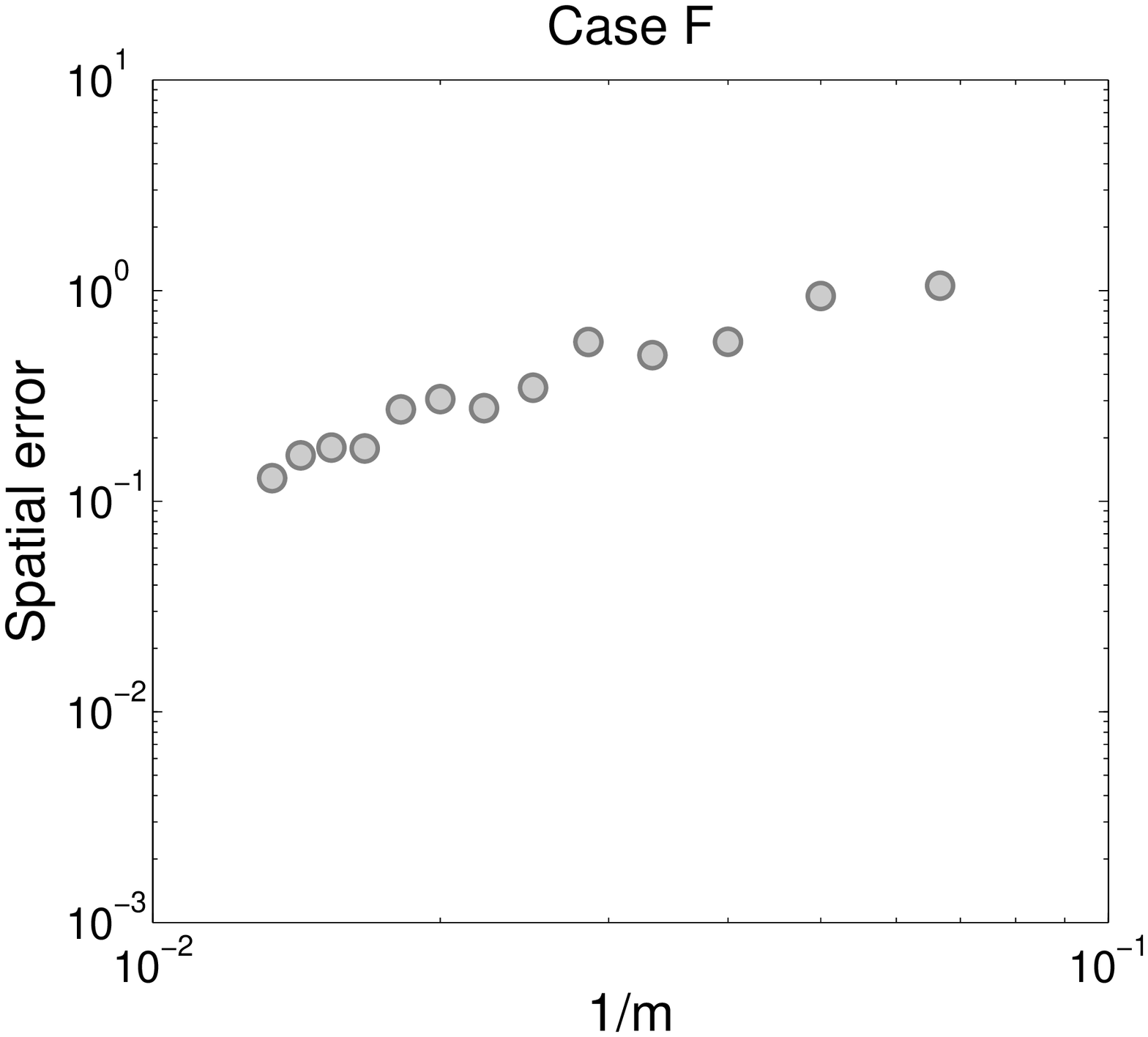}
\end{tabular}
\end{center}
\caption{Spatial discretization errors with uniform $2m\times m\times m$ grid
for European call options in the six cases of Table~\ref{cases1} with
$\rho_{13}=\rho_{23}=0$ for $m=10,15,\ldots,75$. \vspace{9mm} }
\label{SpatialError0}
\end{figure}

We first consider the FD discretization and study the spatial
discretization errors defined by (\ref{es}), with region of interest
\begin{equation*}
{\cal D}=(\tfrac{1}{2}K,\tfrac{3}{2}K)\times (0,1)\times (0,\tfrac{1}{4}).
\end{equation*}
As mentioned above, it is possible to compute these whenever
$\rho_{13}=\rho_{23}=0$.
Figure~\ref{SpatialError1} displays the errors $e(2m,m,m)$ vs $1/m$ in
the six pertinent cases of Table~\ref{cases1} for $m=10,15,\ldots,75$.
Note that $m=75$ means $M=855000$ spatial grid points, which was the
practical (memory) limit on our laptop computer.
Figure~\ref{SpatialError1} clearly shows that in each case the spatial
discretization errors decrease as $m$ increases.
To determine the numerical orders of convergence, straight lines have
been fitted to the results.
In the cases A, B, C, D, F the obtained orders of convergence are all
equal to two approximately.
Only in case E a slightly lower order was obtained, namely 1.6.
As an indication of the sizes of the spatial discretization errors in
a relative sense, we mention that these always lie between 0.2\% and
1.2\% when $m=50$ and between 0.1\% and 0.6\% when $m=75$ (here only
option values are considered that are greater than~1).
In view of the foregoing, we conclude that the FD discretization defined
in Section~\ref{space} performs satisfactory in all six cases.
It is interesting to briefly compare the spatial errors to those obtained
with a uniform grid and the same number of grid points.
Figure~\ref{SpatialError0} shows spatial discretization errors analogously
to Figure~\ref{SpatialError1}, but then for uniform $2m\times m\times m$
grids.
Clearly, in most cases the spatial errors for the nonuniform grid are
substantially smaller, often by an order of magnitude, than those for the
corresponding uniform grid.
Further, it is clear that for a uniform grid the behavior of the spatial
error as a function of the number of grid points is erratic, which is
undesirable.
Also, the nonuniform grid yields more points in the region in
$(s,v,r)$--space where one wishes to obtain option prices.
We therefore conclude that the nonuniform grid defined in Section
\ref{space} is preferable over a uniform grid.

\begin{figure}
\begin{center}
\begin{tabular}{c c}
         \includegraphics[width=0.5\textwidth]{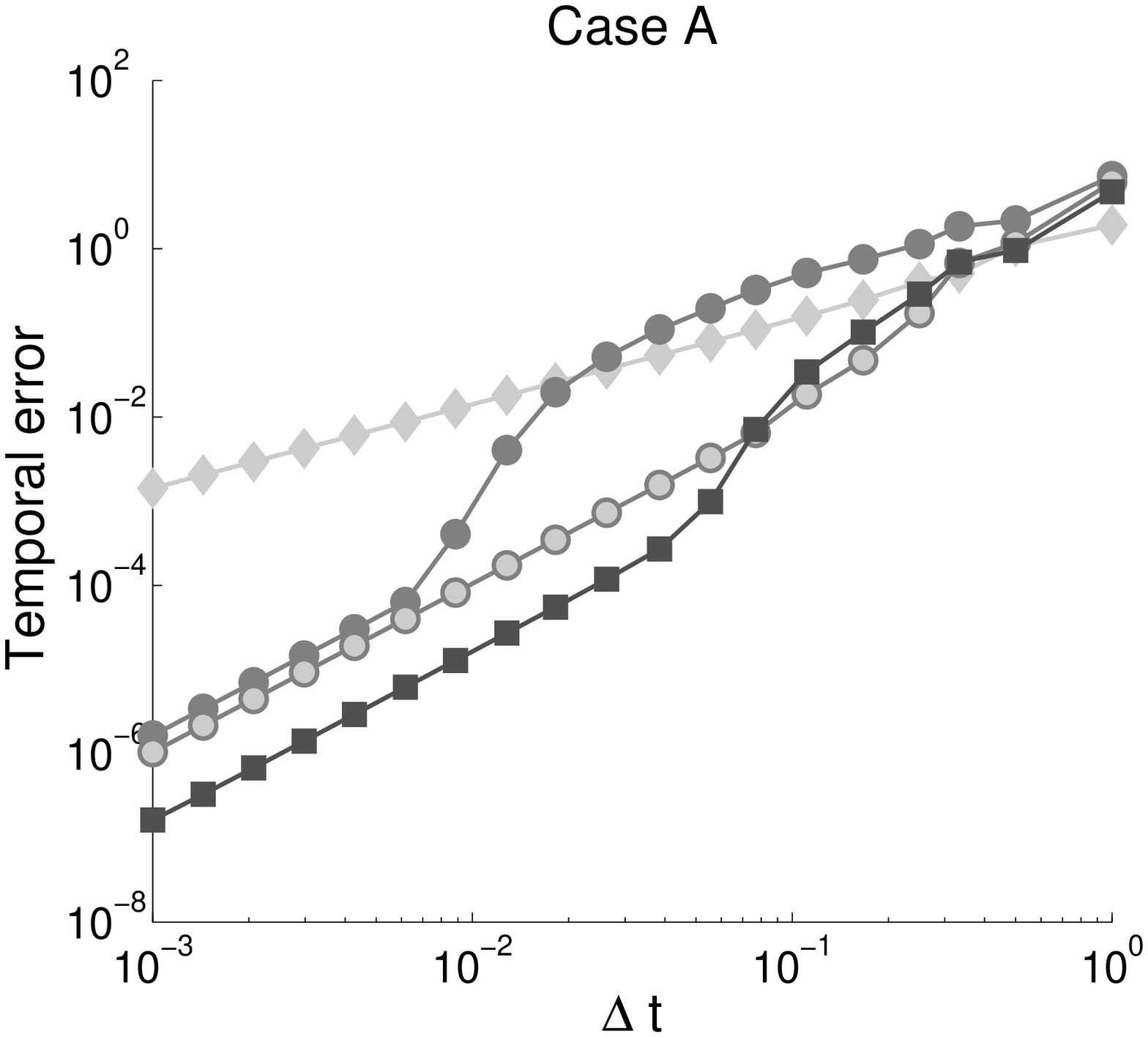}&
         \includegraphics[width=0.5\textwidth]{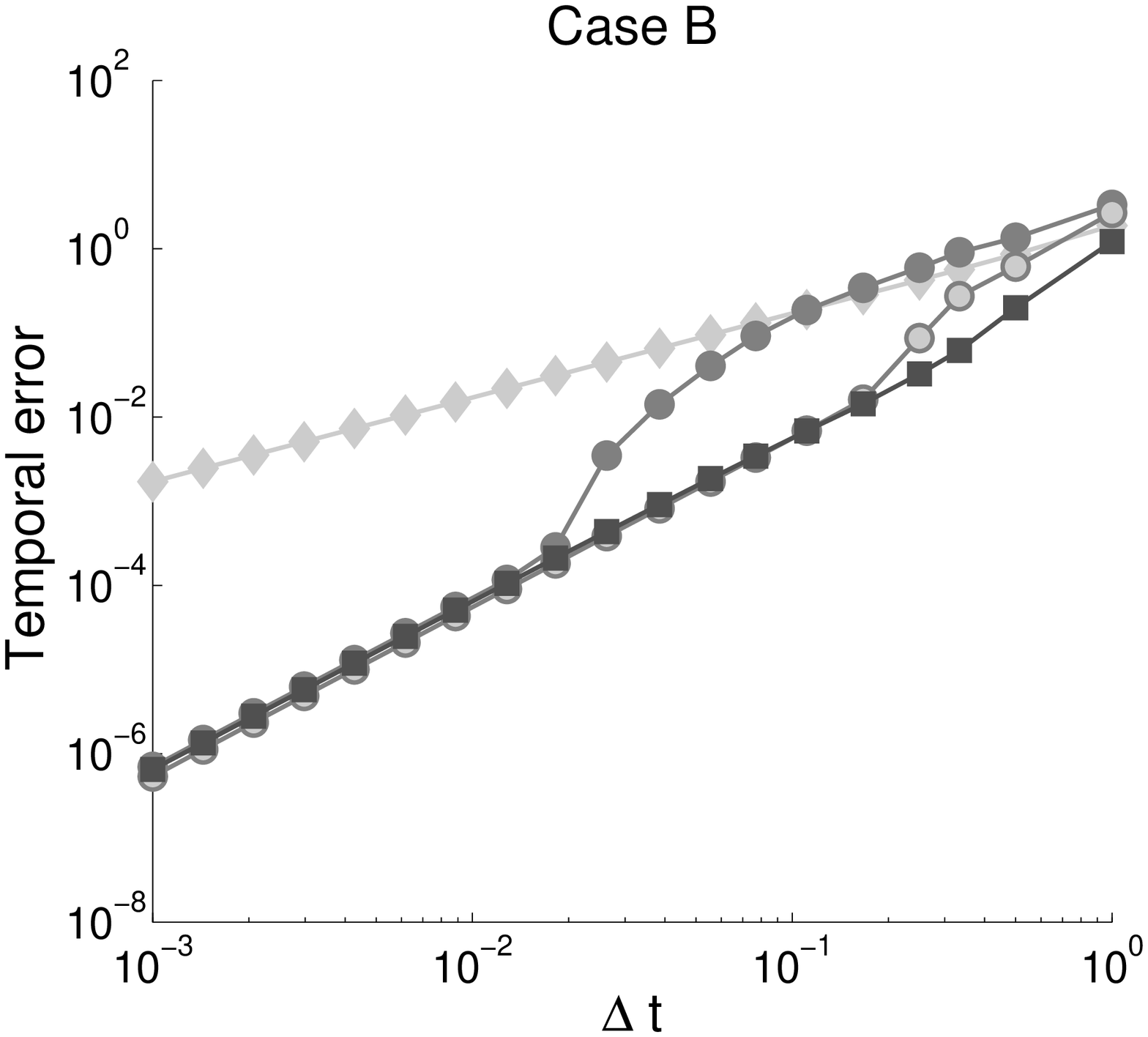}\\
         \includegraphics[width=0.5\textwidth]{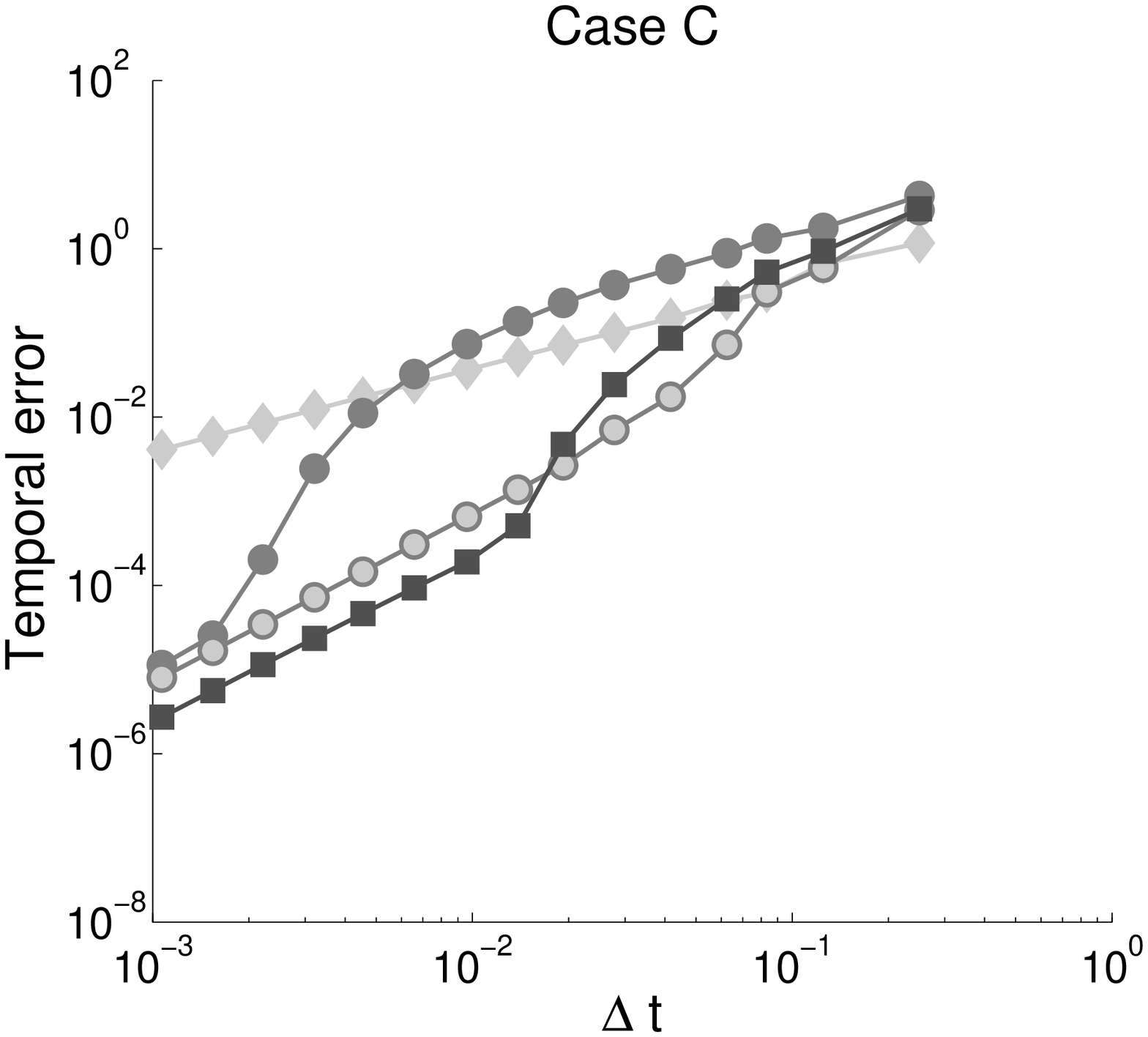}&
         \includegraphics[width=0.5\textwidth]{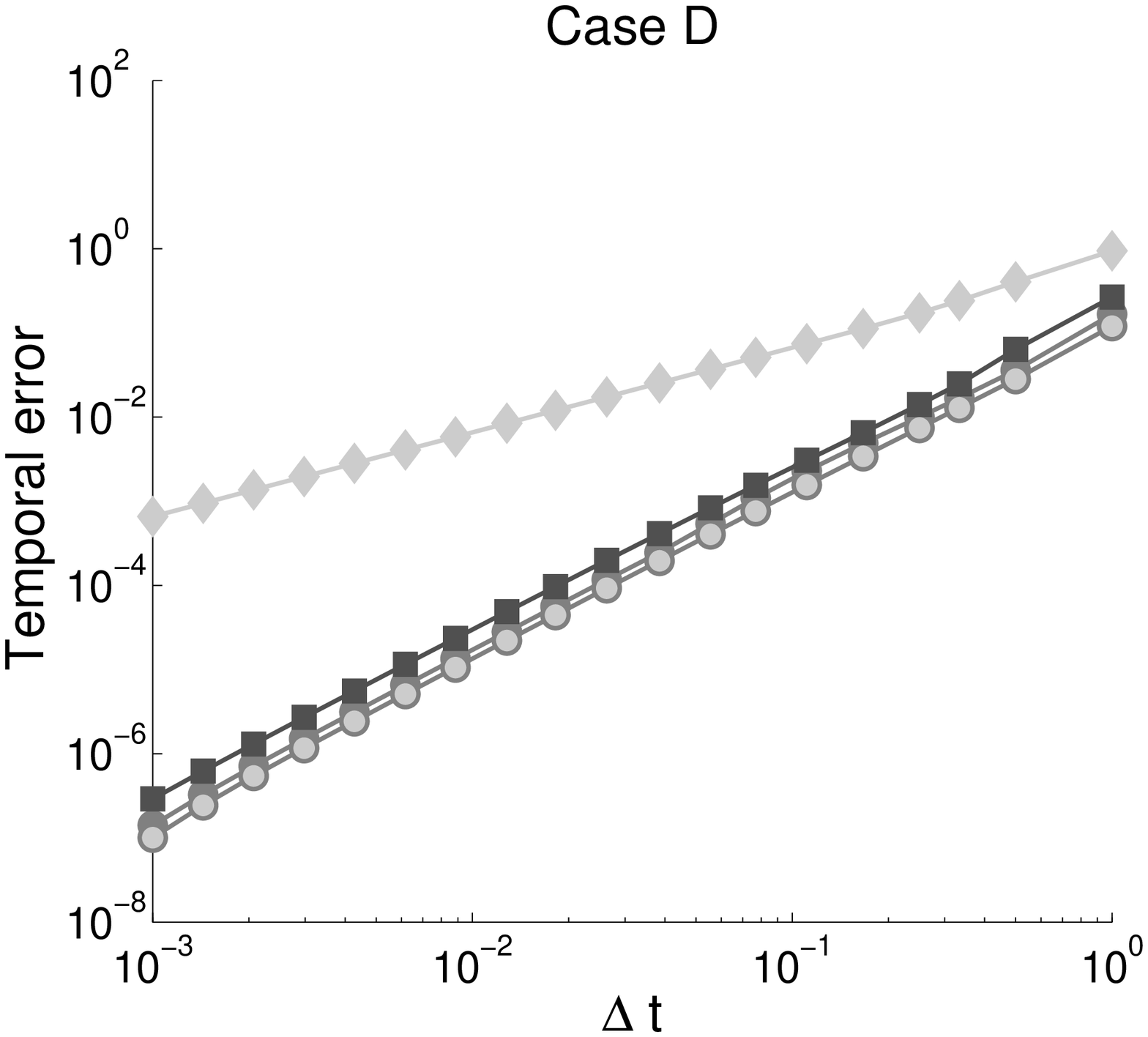}\\
         \includegraphics[width=0.5\textwidth]{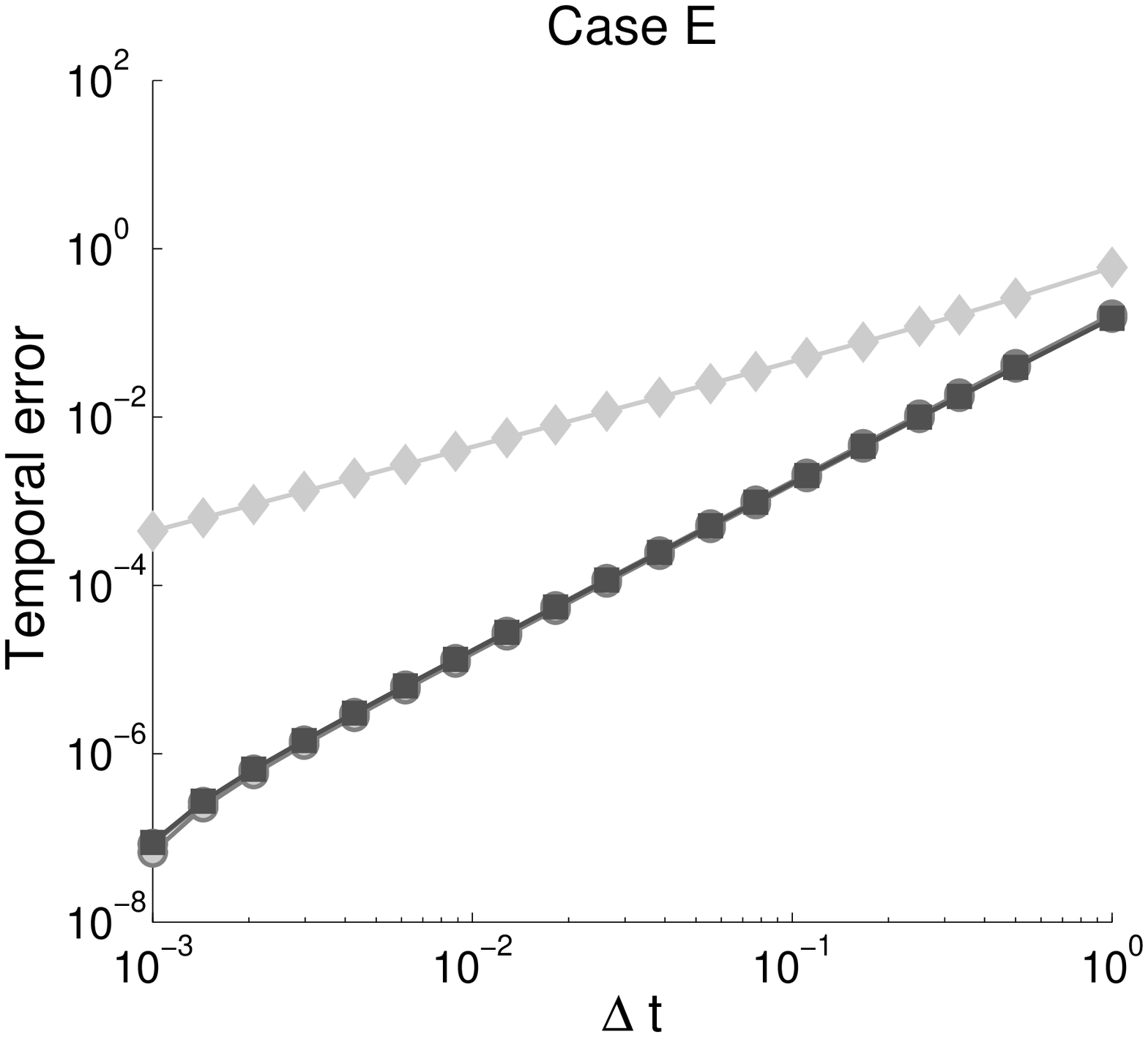}&
         \includegraphics[width=0.5\textwidth]{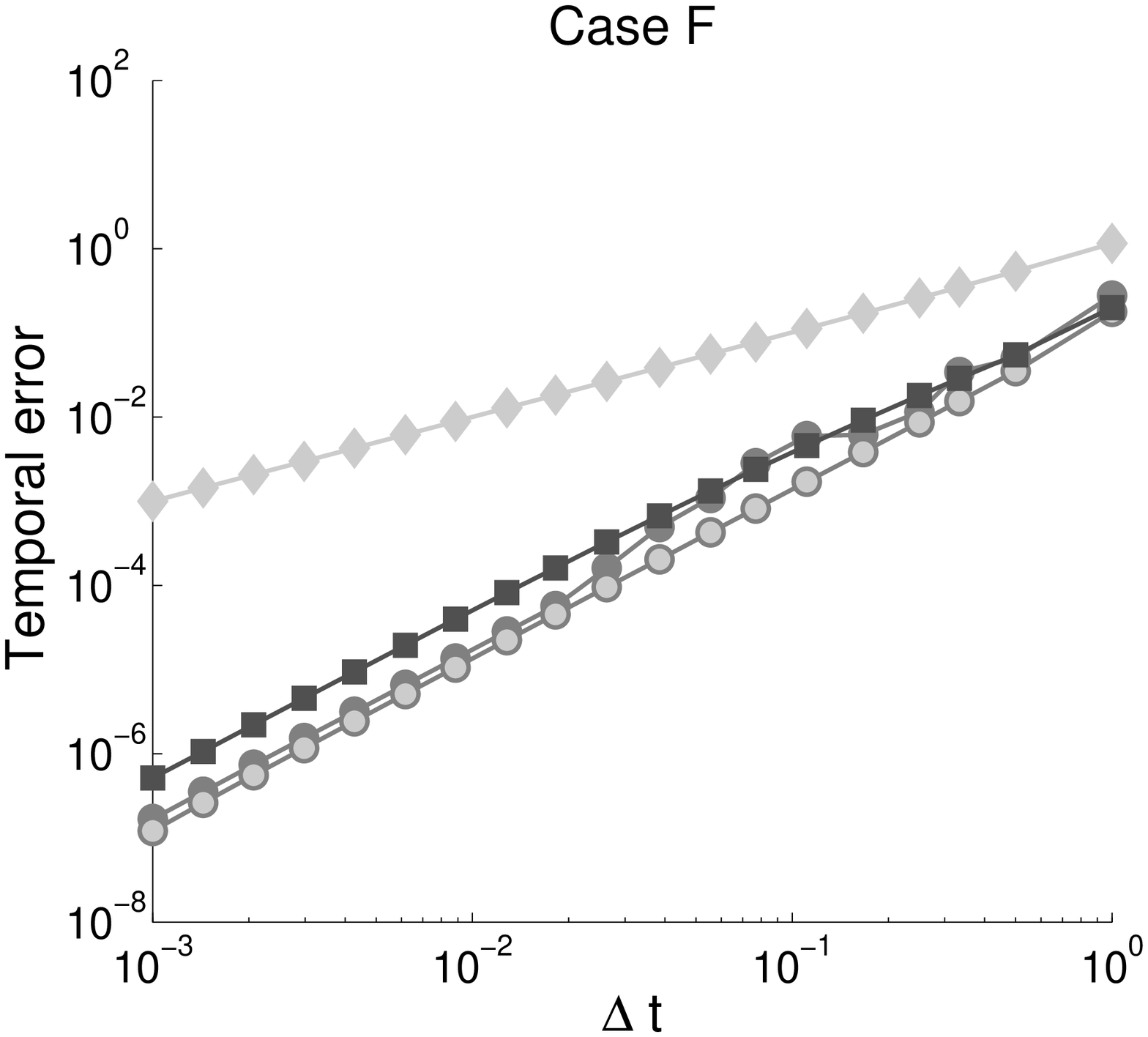}
\end{tabular}
\end{center}
\caption{Temporal discretization errors $\widehat{e}\,(\Delta t;100,50,50)$
vs $\Delta t$ for European call options in the six cases of Table~\ref{cases1}.
ADI schemes: Do with $\theta=\frac{2}{3}$ (diamond), CS with $\theta=\frac{1}{2}$
(dark circle), MCS with $\theta=\max\{\frac{1}{3},\frac{2}{13}(2\gamma+1)\}$
(light circle) and HV with $\theta=\frac{1}{2}+\frac{1}{6}\sqrt{3}$ (square).}
\label{TemporalError1}
\end{figure}

We next consider the performance of the four ADI schemes in the
application to the semidiscrete HHW PDE for European call options
in the six cases of Table~\ref{cases1} with all correlations nonzero.
Figure~\ref{TemporalError1} displays the temporal discretization
errors $\widehat{e}\,(\Delta t;2m,m,m)$ for a sequence of step sizes
with $10^{-3} \le \Delta t \le 10^0$ when $m=50$.

A first main observation from Figure~\ref{TemporalError1} is that for
all four ADI schemes the temporal discretization errors are bounded
from above by a moderate value and decay monotonically as $\Delta t$
decreases.
Additional experiments indicate that this is true for any value $m$;
see a further discussion below.
This suggests an unconditionally stable behavior of the schemes, 
which is a new and nontrivial result.
It does not directly follow for example from the von Neumann
stability analysis presented in Section~\ref{ADI}.
We note that this result holds in all six cases, independently
of whether or not the Feller condition is fulfilled.

A next observation is that the CS scheme exhibits an undesirable feature
in the cases \mbox{A, B, C} with temporal errors that are very large for
moderate $\Delta t$, compared to what may be expected on the basis of its
asymptotic error behavior (ie, for small $\Delta t$).
To a much lesser extent, this is also observed with the HV and MCS
schemes.
Additional experiments reveal that the relatively large temporal
errors occur at spatial grid points near the strike $K$.
It is already known in the literature that the nonsmoothness
of the initial function at the strike yields high-frequency errors
which are not always sufficiently damped by numerical schemes,
notably the Crank--Nicolson scheme and the Do and CS schemes with
$\theta = \tfrac{1}{2}$.
A popular remedy for this situation is to first apply, at $t=0$, two
implicit Euler steps with step size $\Delta t /2$, and then to proceed
onwards from $t=\Delta t$ with the scheme under consideration, 
cf Rannacher (1984).
However, in our present application of the three-dimensional HHW PDE
this damping procedure is computationally intensive.
We shall consider an alternative in the next subsection.

A further analysis of the results in Figure~\ref{TemporalError1} indicates
that in each case the temporal discretization errors for the Do scheme  
are bounded from above by $C \Delta t$ and for the MCS, HV schemes by
$C (\Delta t)^2$ (whenever $\Delta t >0$) with constants $C$ depending
on the scheme and the case.
This clearly agrees with the respective orders of consistency of the
schemes.
Moreover, experiments with both smaller and larger values of $m$ suggest 
that the constants $C$ are only weakly dependent on the number of spatial 
grid points $M$, ie, the error bounds are valid in a stiff sense, which 
is a desirable property.
This result is also nontrivial, as the order of consistency is a priori
only relevant to fixed, nonstiff systems of ODEs.
For the CS scheme, we find that the temporal errors can be bounded in
each case by $C (\Delta t)^2$ with a constant $C$ independent of 
stiffness if damping is applied.
Actual numerical experiments for ADI schemes combined with damping 
will be presented in the next subsection.

Our implementation of the ADI finite difference discretization has been
done in Matlab, where all matrices have been defined as sparse.
For the CS, MCS, HV schemes the cpu-time per time step was about 0.10,
0.18, 0.90, 1.5 cpu-seconds for $m =$ 25, 30, 50, 60, respectively,
on one Intel Core Duo T7250 2.00 GHz processor with 4 GB memory;
for the Do scheme these times are about halved.
Here all correlations were nonzero and the mean reversion level
was time-dependent.
It readily follows that the cpu-times are indeed almost directly
proportional to the number of spatial grid points $M \sim 2m^3$.

\subsection{Up-and-out call options}

As an important and particularly challenging type of exotic options we
consider here European-style up-and-out call options.
The FD discretization described in Section~\ref{space} is adapted with 
few modifications.
Let barrier $S_{\max}=:B>K$ be given.
Then the boundary conditions (\ref{BC}b), (\ref{BC}c) are replaced by
\begin{subeqnarray}\label{BC2}
\phantom{\frac{\partial u}{\partial s}}u(s,v,r,t)&=~~0
\quad &{\rm whenever}~~s=B,\\
\frac{\partial u}{\partial v}(s,v,r,t)&=~~0
\quad &{\rm whenever}~~v=V_{\max}.
\end{subeqnarray}
\vskip0.1cm\noindent
The condition (\ref{BC2}b) has been suggested by various authors in the
literature.
Note that all boundary conditions are now homogeneous, and $g(t) \equiv 0$.
The relevant set of spatial grid points is
\begin{align*}
\mathcal{G} = \{(s_i, v_j, r_k): 1\leq i \leq m_1-1\,,\, 0\leq j\leq m_2
\,,\, 0\leq k\leq m_3\}.
\end{align*}
The only significant change we make to the FD discretization of
Section~\ref{space} is to replace, in the $s$-direction, the central
advection scheme \eqref{MethodB} by the backward scheme (\ref{backw})
if $r<0$ and by the forward scheme (\ref{forw}) if $r>0$.
This upwind approach alleviates spurious oscillations in the FD 
solution that are obtained with the central advection scheme.
It is already useful for up-and-out call options in the 
one-dimensional Black--Scholes model.
The pricing of up-and-out call options is numerically more challenging 
than of vanilla options, due to the boundary layer that is introduced 
at the barrier.

\begin{figure}
\begin{center}
\begin{tabular}{c c}
         \includegraphics[width=0.5\textwidth]{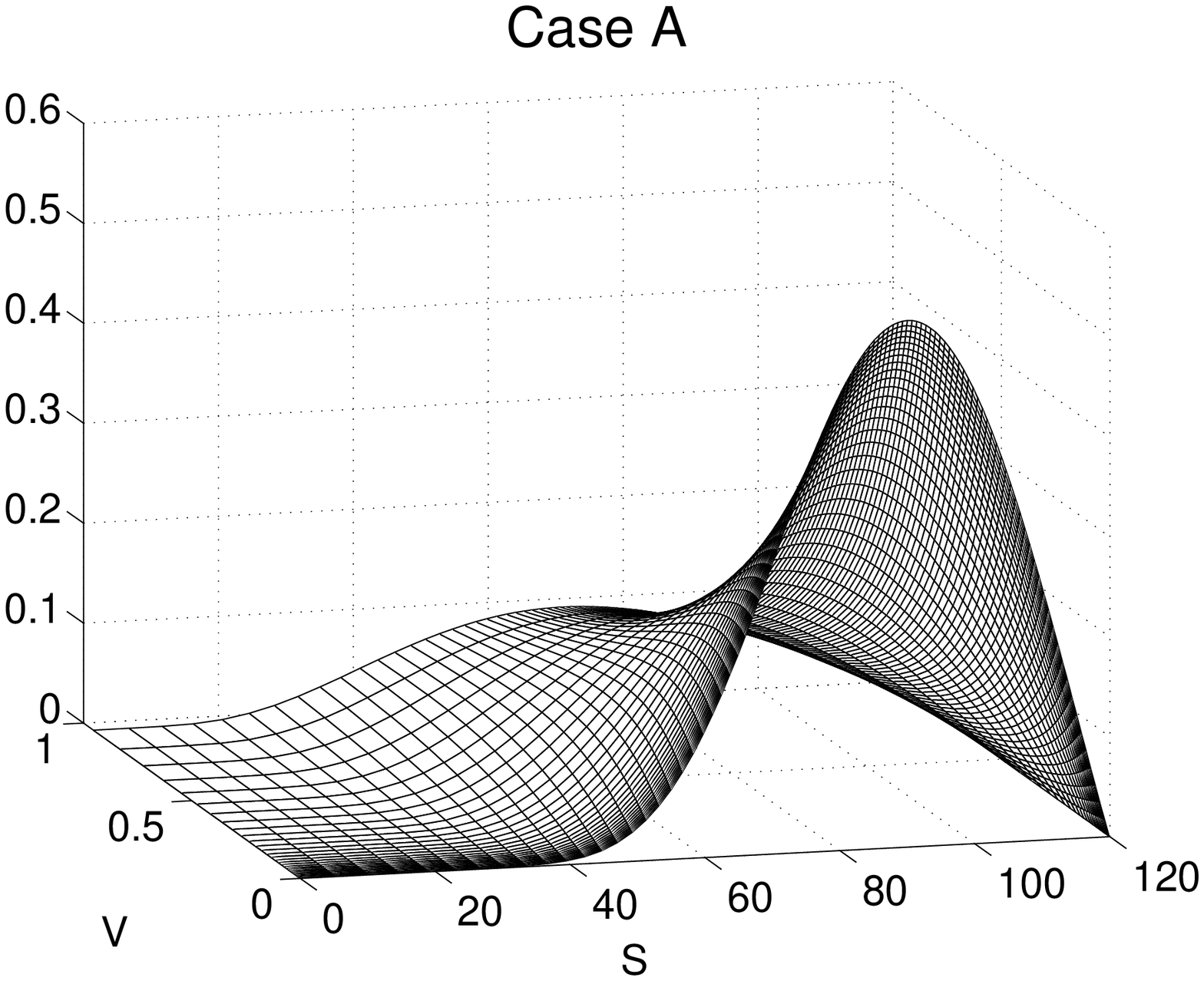}&
         \includegraphics[width=0.5\textwidth]{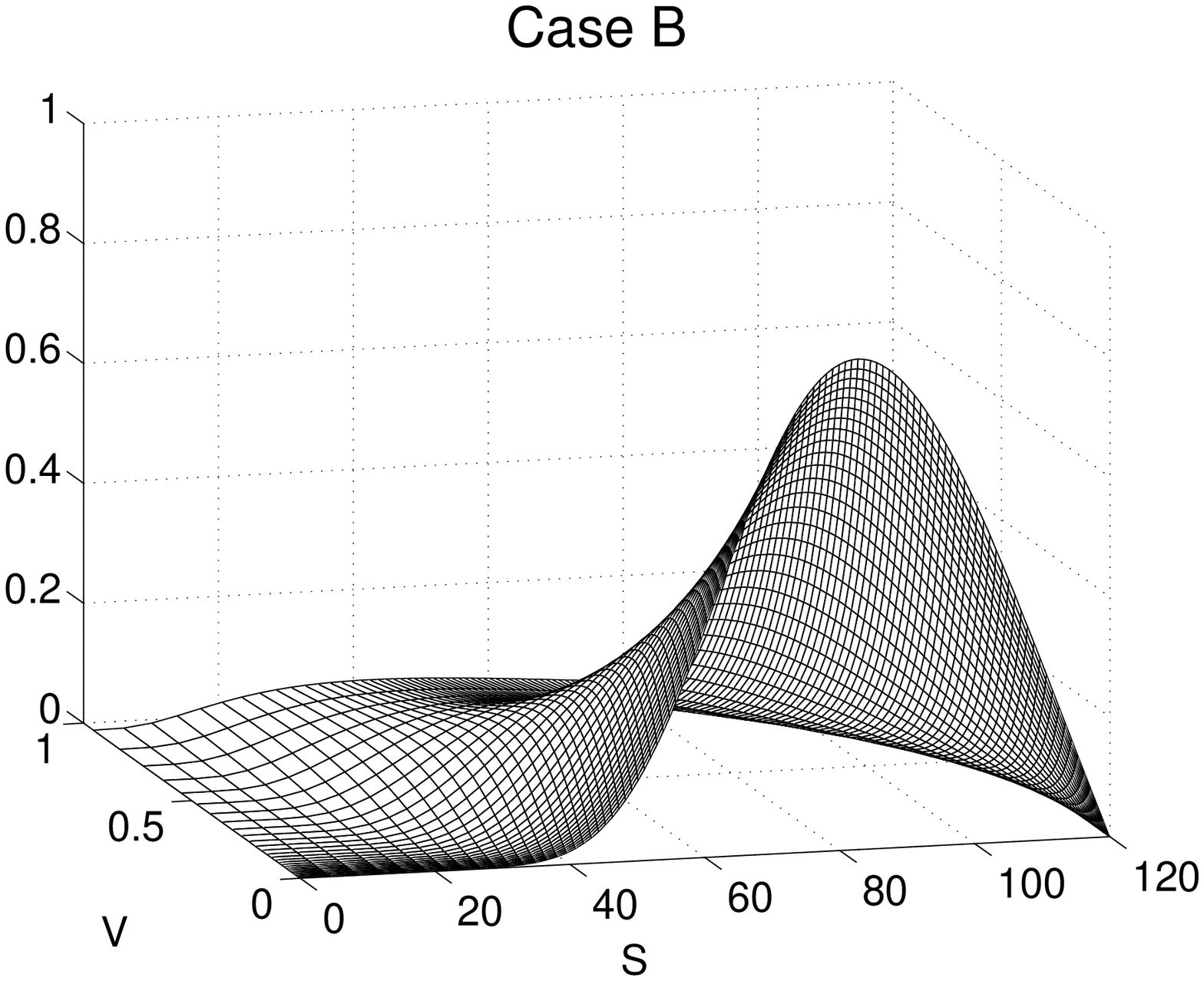}\\
         \includegraphics[width=0.5\textwidth]{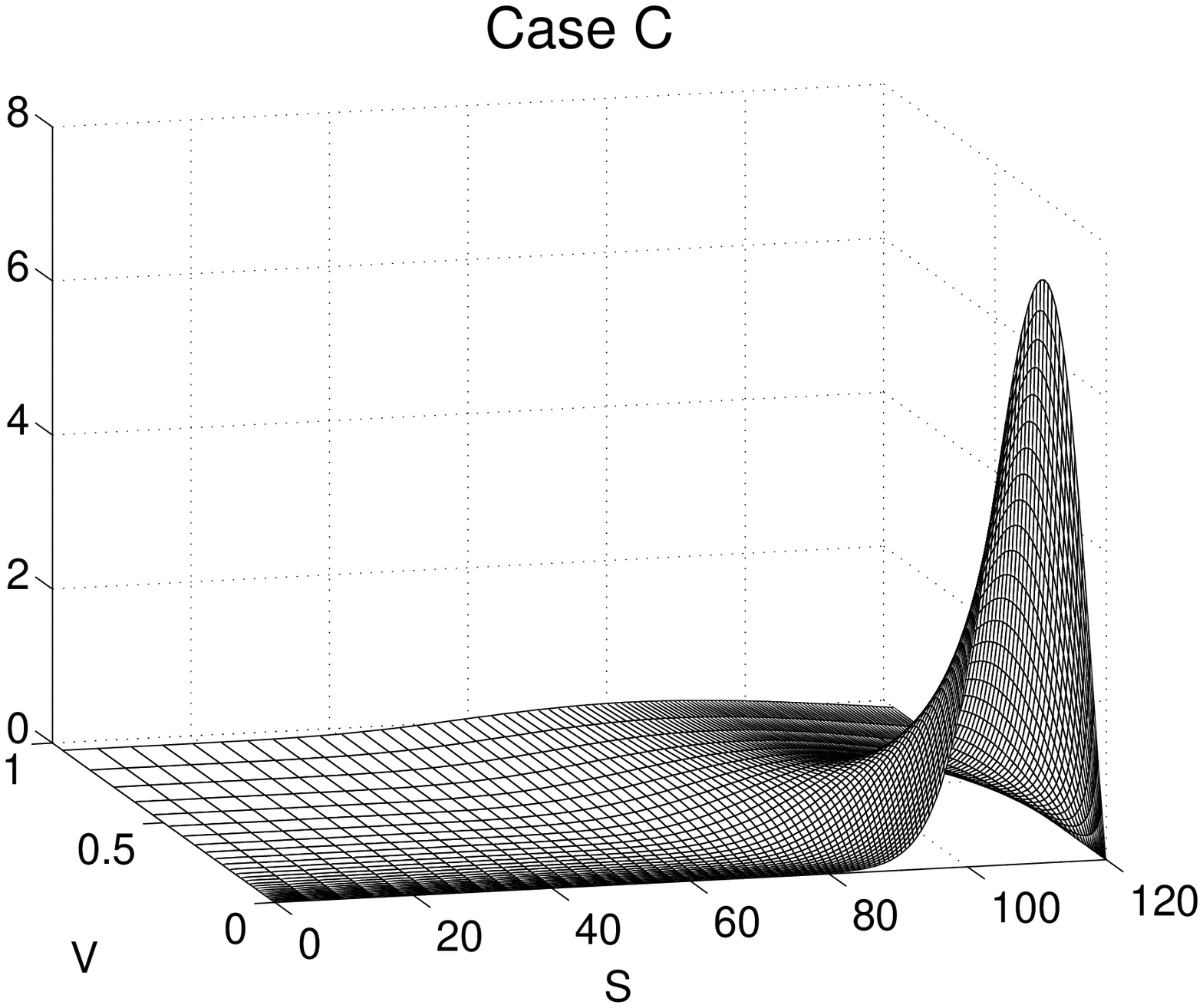}&
         \includegraphics[width=0.5\textwidth]{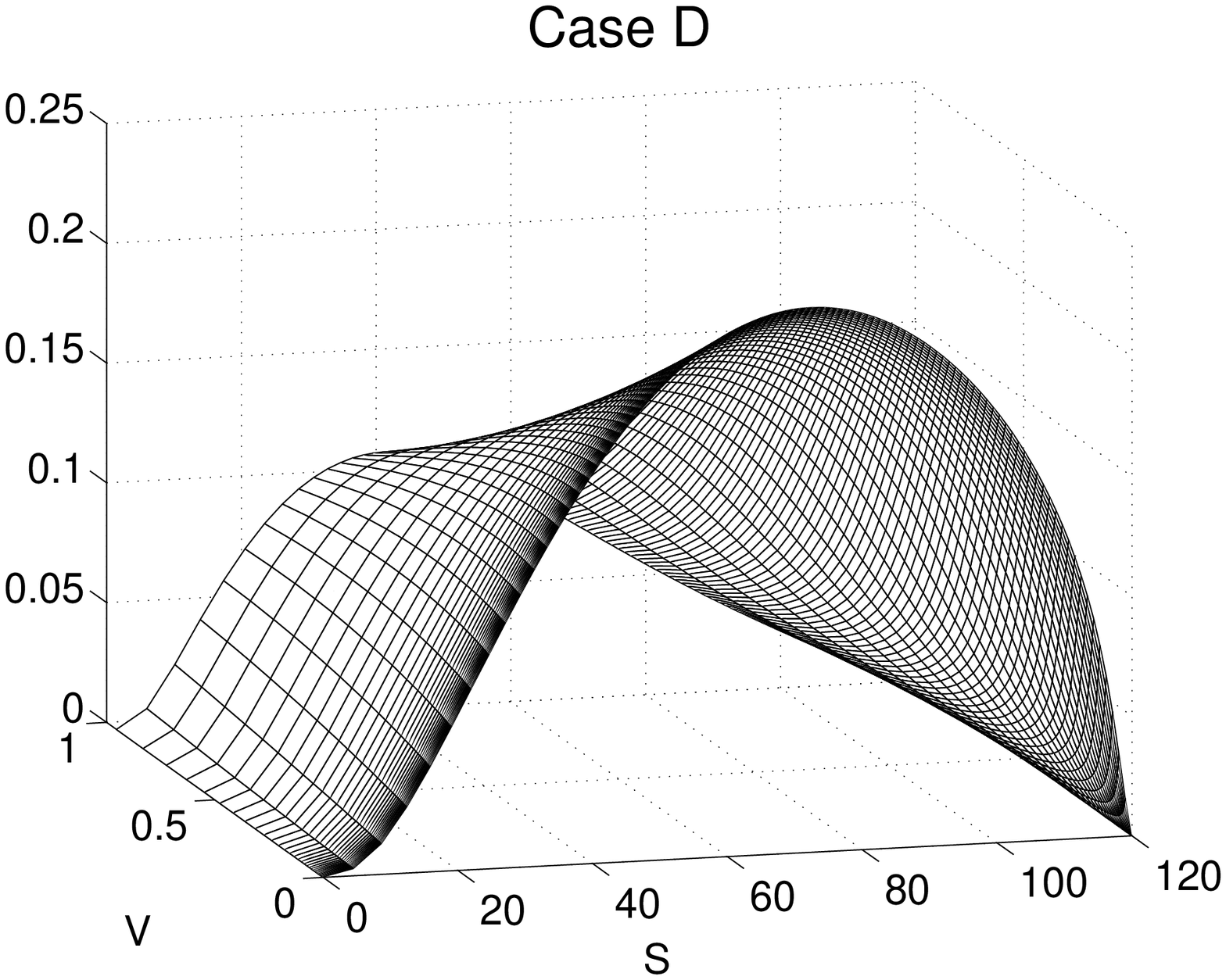}\\
         \includegraphics[width=0.5\textwidth]{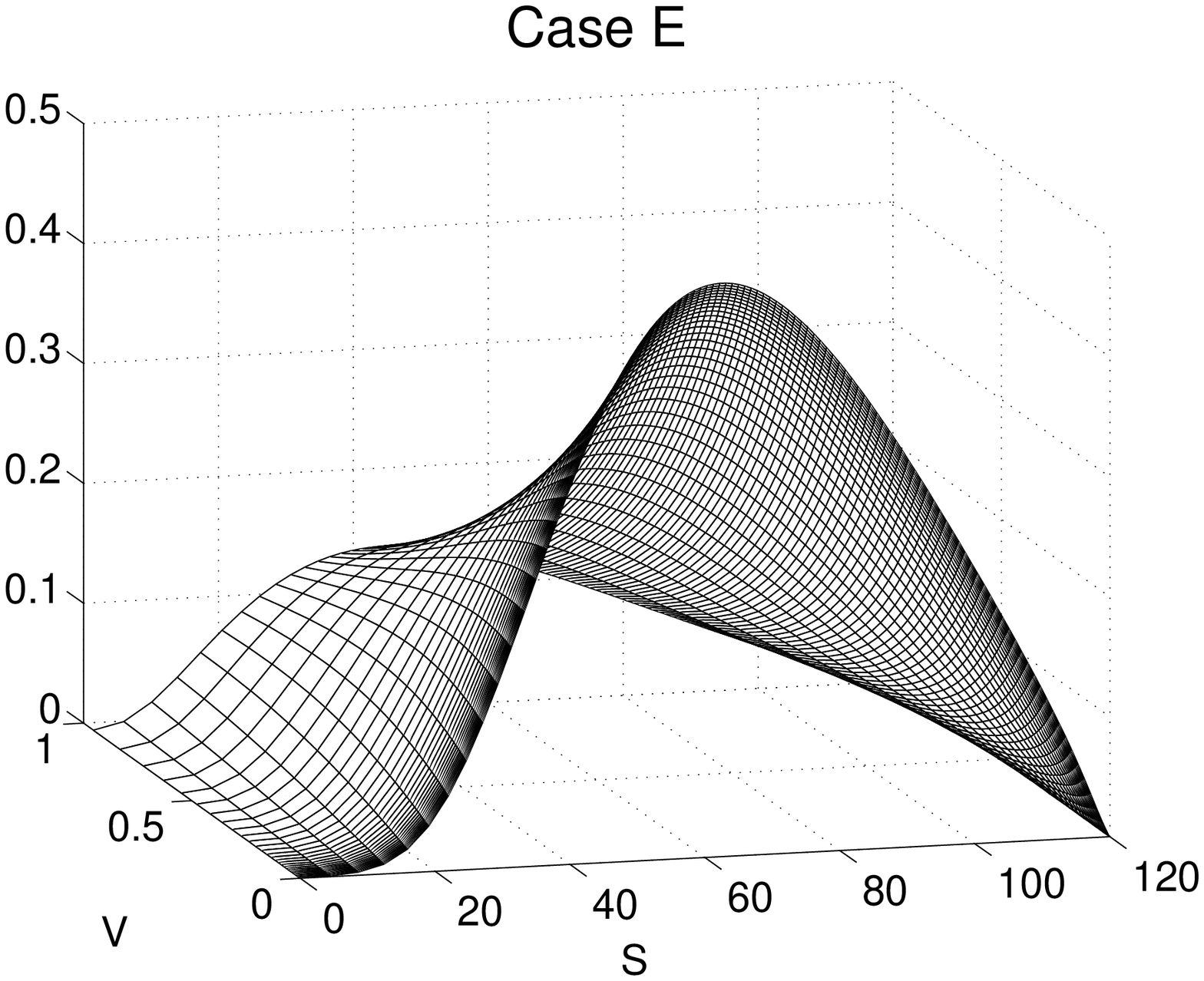}&
         \includegraphics[width=0.5\textwidth]{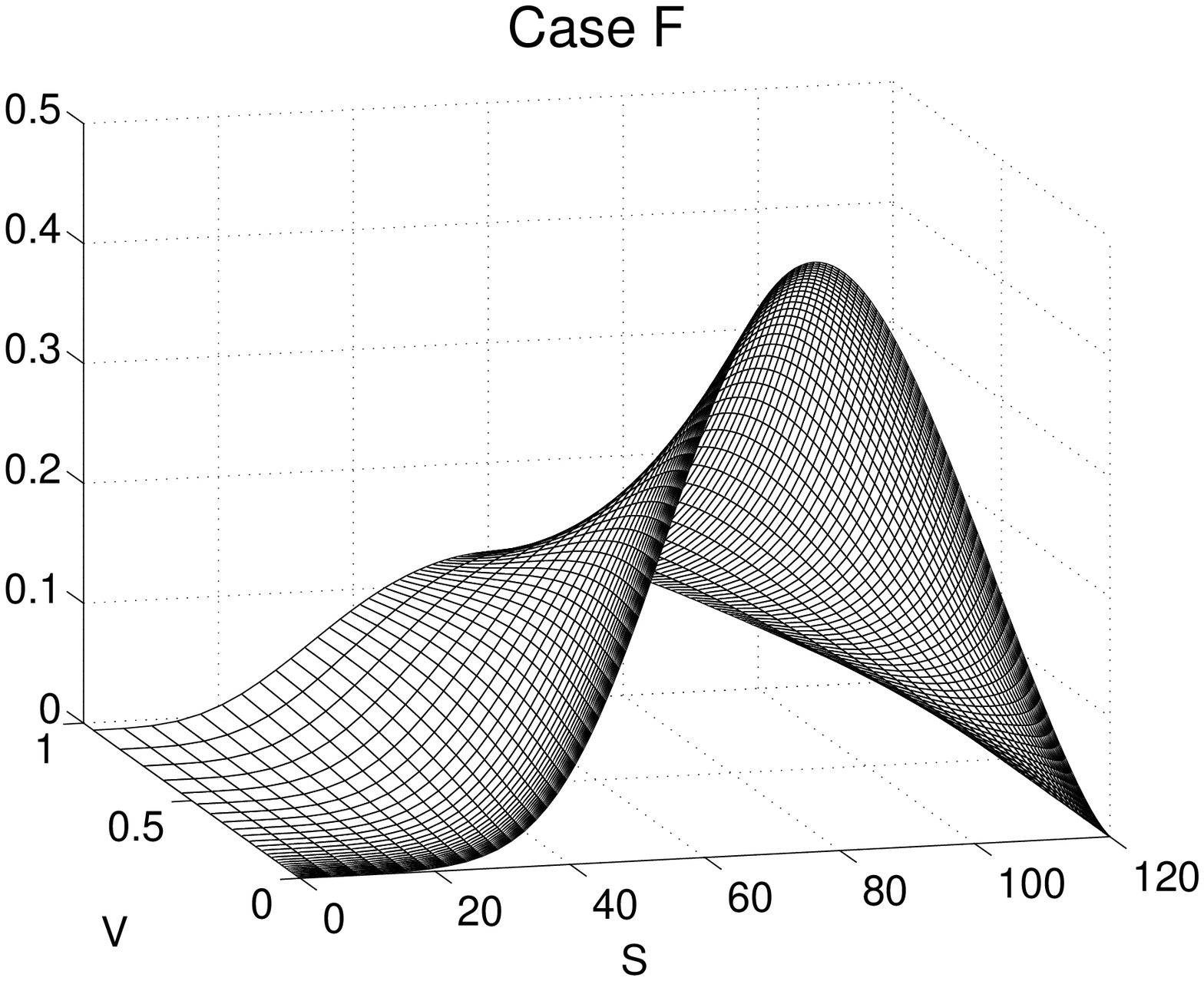}
\end{tabular}
\end{center}
\caption{European up-and-out call option values in all cases of 
Table~\ref{cases1} with barrier $B=120$.
Spot interest rates in the cases A, B, C, D, E, F are, respectively,
$r=$ 0.025, 0.022, 0.025, 0.027, 0.022, 0.017.
Note: scales on vertical axes vary.}
\label{ExactBarrier}
\end{figure}

\begin{figure}
\begin{center}
\begin{tabular}{c c}
         \includegraphics[width=0.5\textwidth]{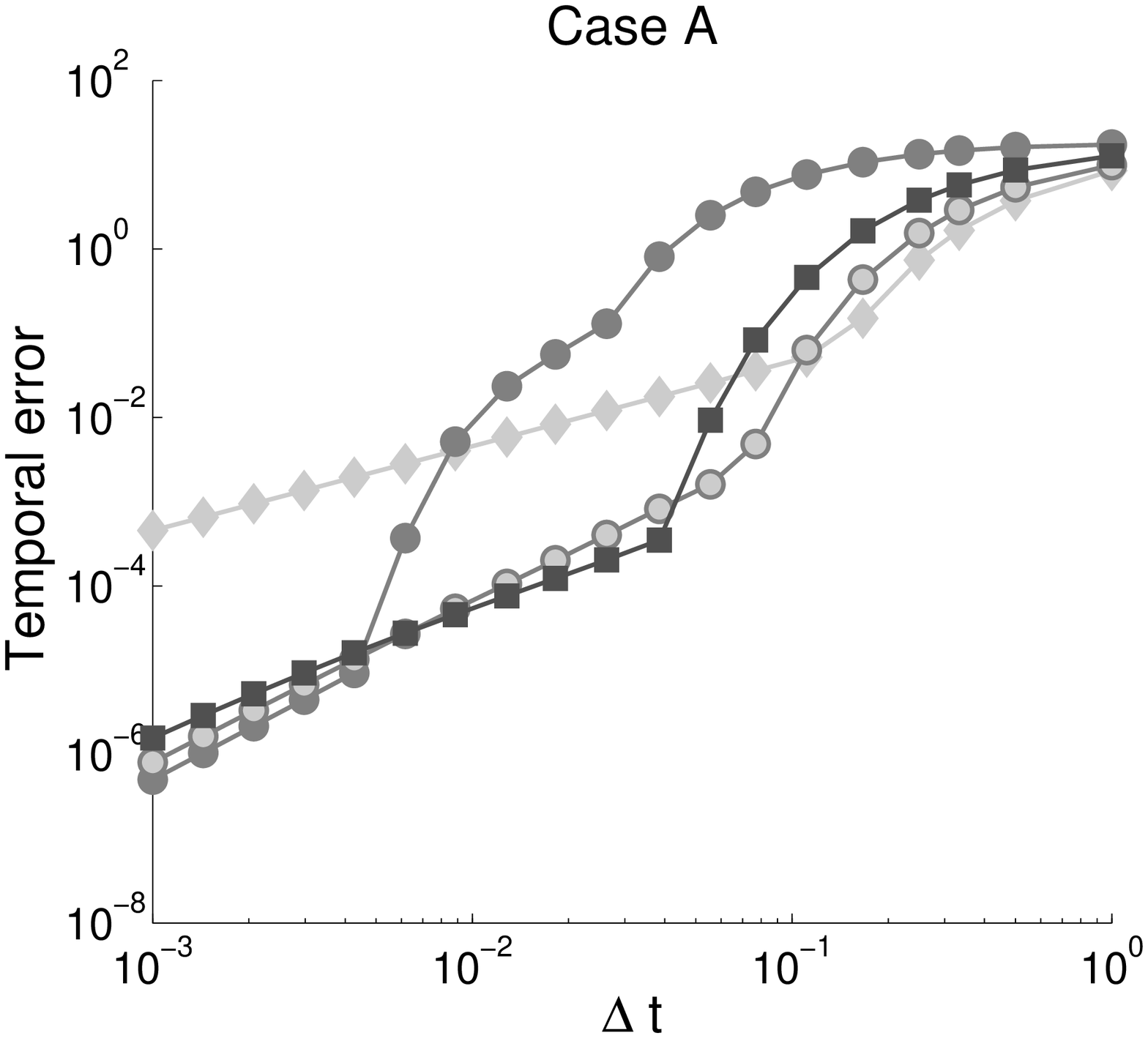}&
         \includegraphics[width=0.5\textwidth]{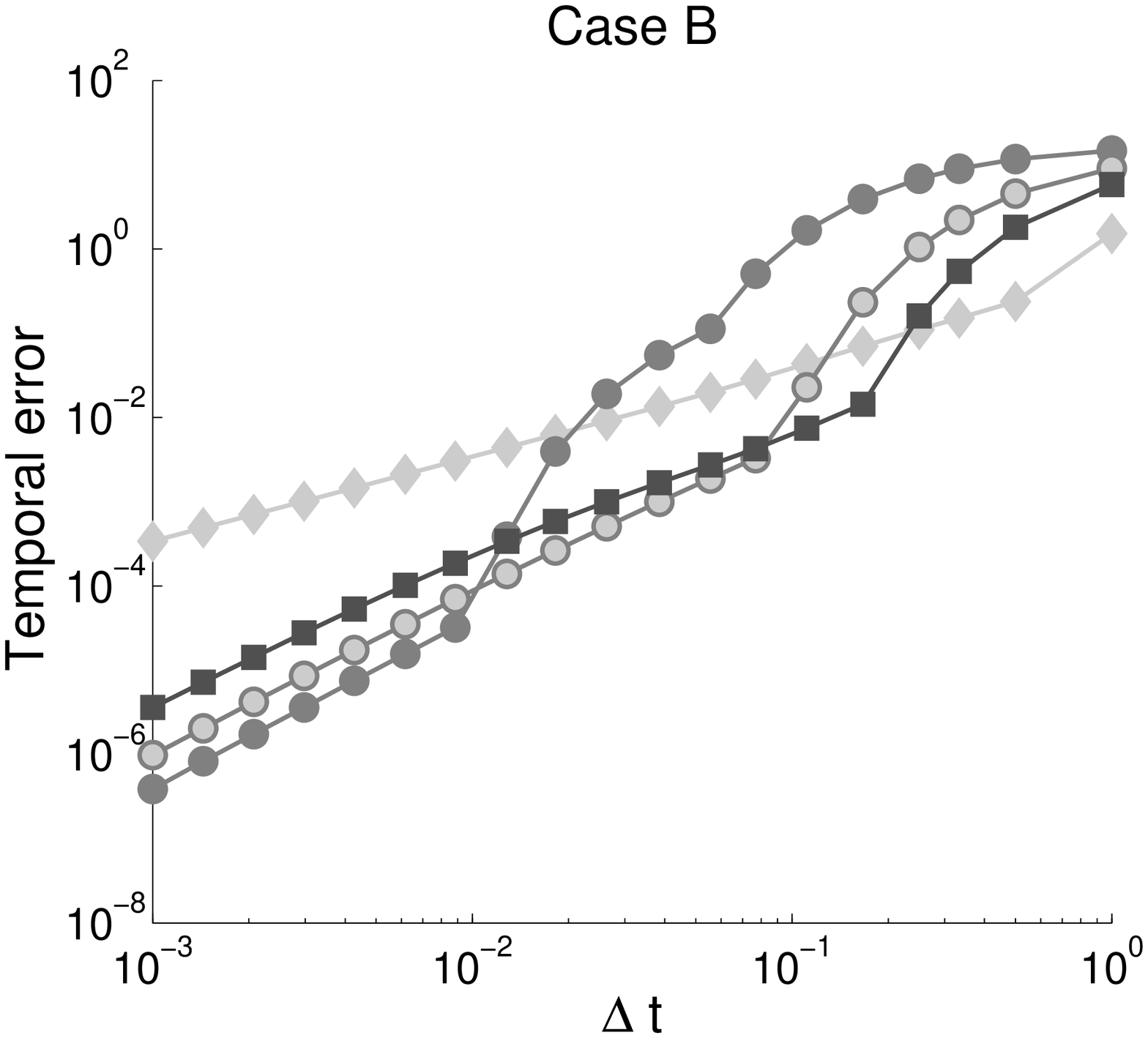}\\
         \includegraphics[width=0.5\textwidth]{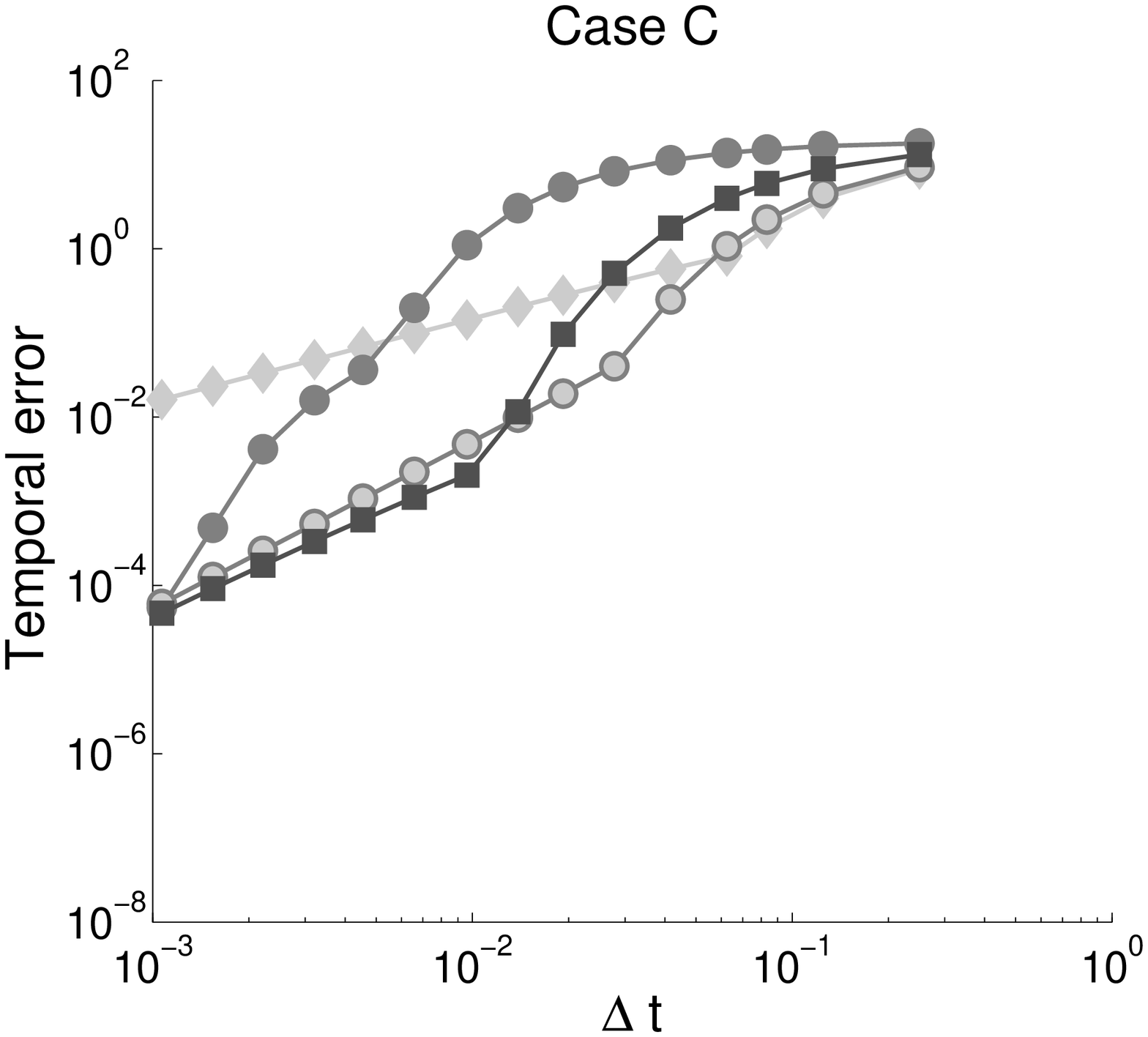}&
         \includegraphics[width=0.5\textwidth]{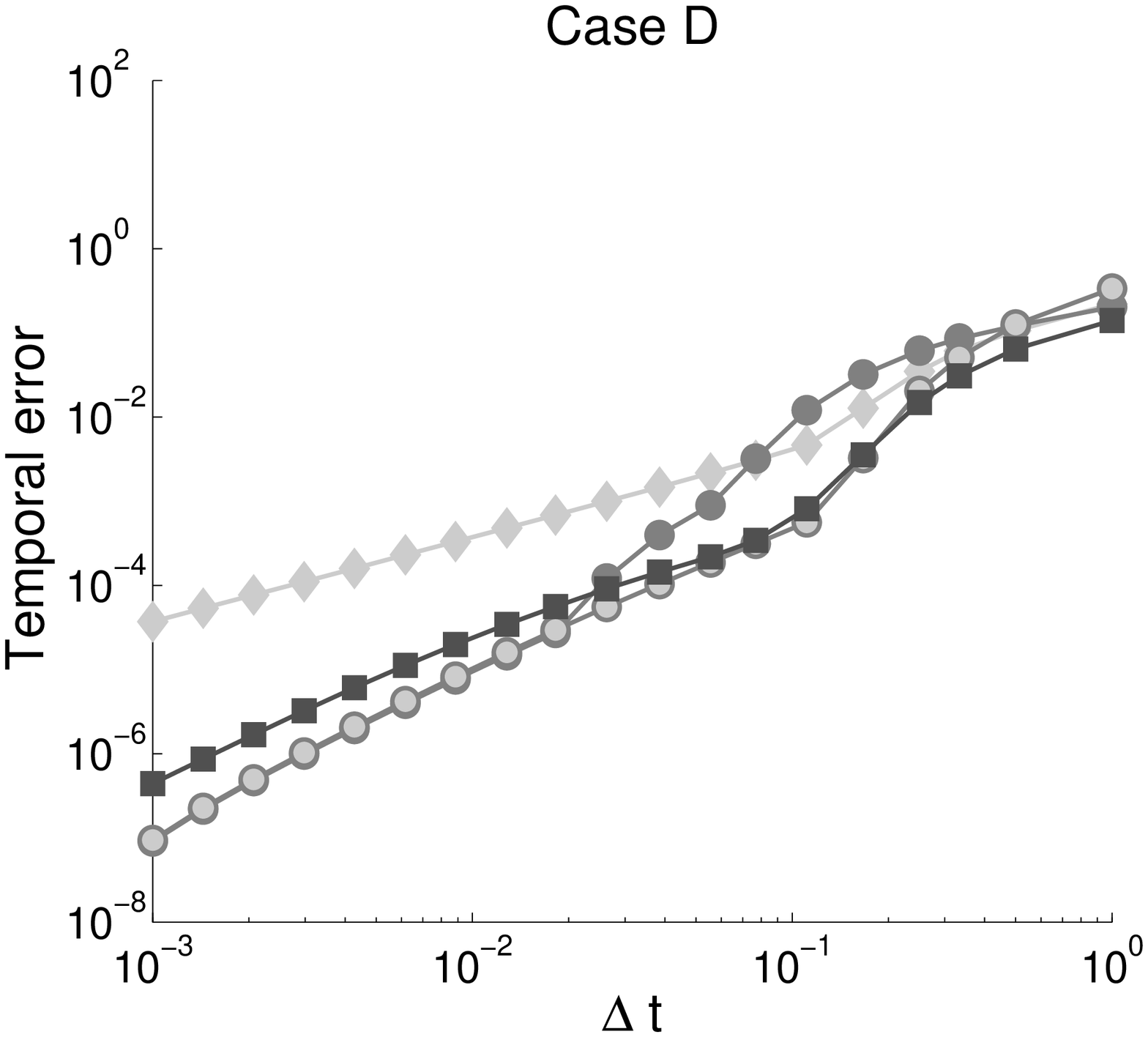}\\
         \includegraphics[width=0.5\textwidth]{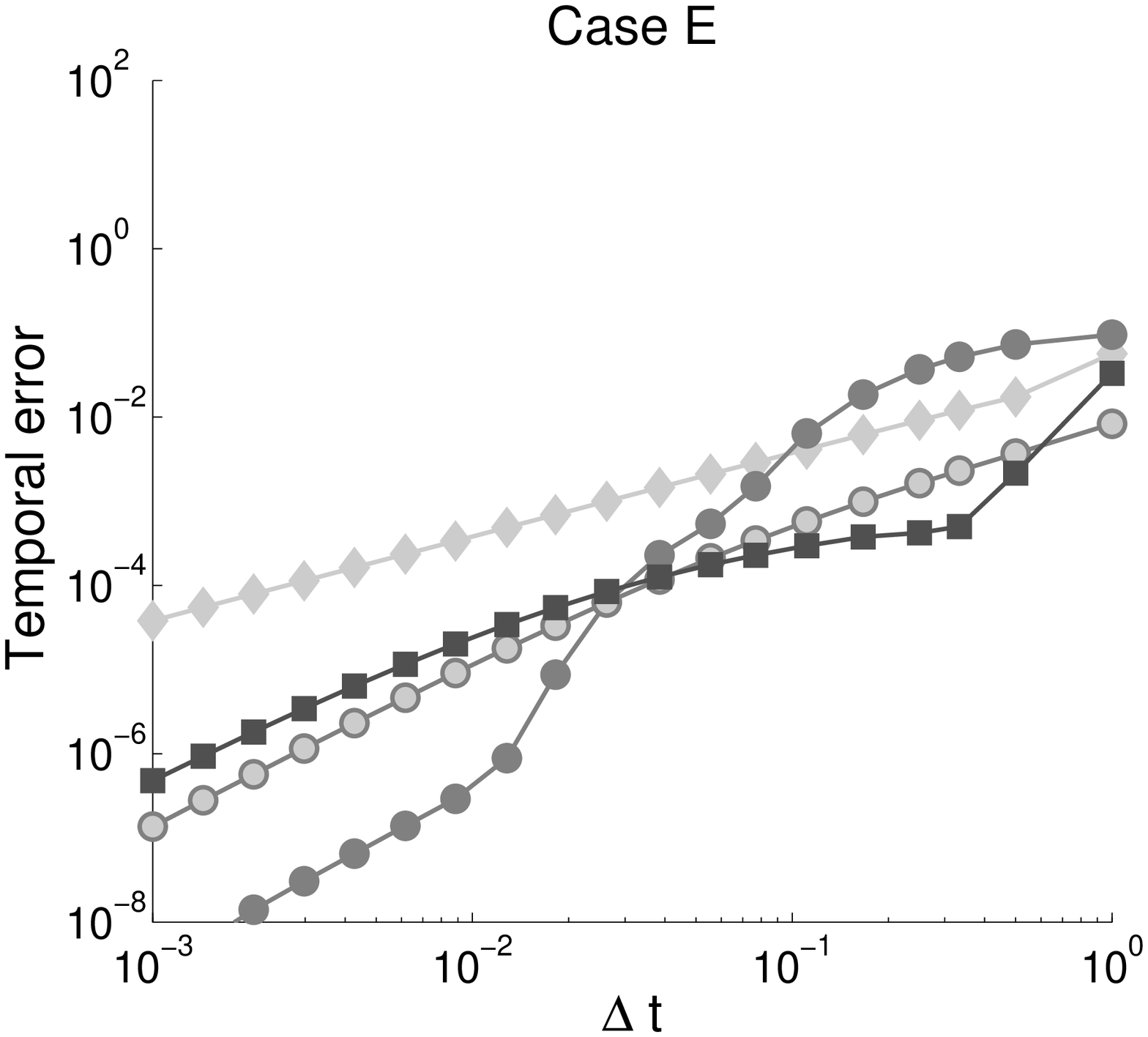}&
         \includegraphics[width=0.5\textwidth]{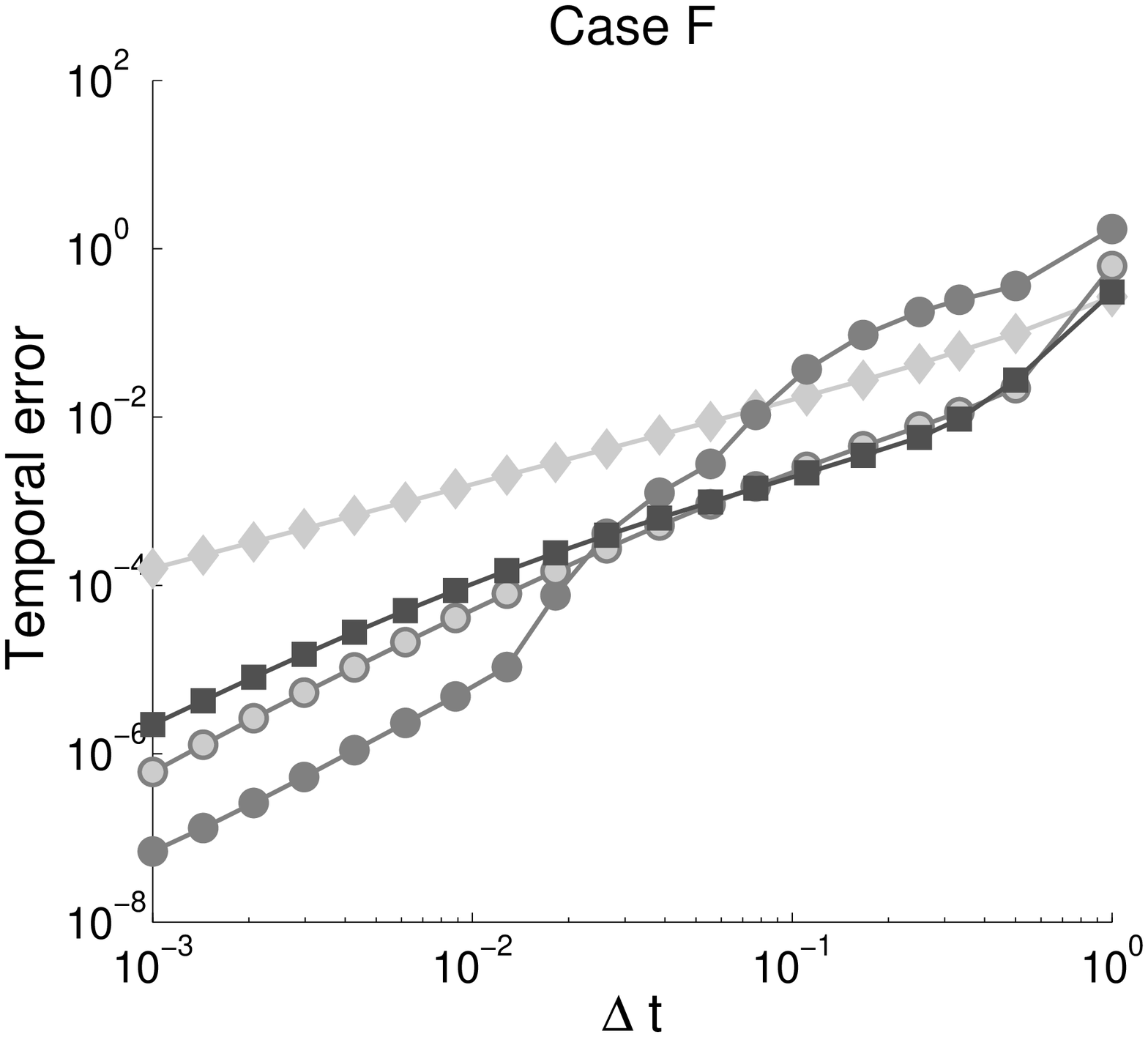}
\end{tabular}
\end{center}
\caption{Temporal discretization errors $\widehat{e}\,(\Delta t;100,50,50)$ vs 
$\Delta t$\, for up-and-out call options in all cases of Table~\ref{cases1} for 
barrier $B=120$.
ADI schemes: Do with $\theta=\frac{2}{3}$ (diamond), CS with $\theta=\frac{1}{2}$
(dark circle), MCS with $\theta=\max\{\frac{1}{3},\frac{2}{13}(2\gamma+1)\}$
(light circle), and HV with $\theta=\frac{1}{2}+\frac{1}{6}\sqrt{3}$ (square). 
~No initial damping.} \label{TemporalError2a}
\end{figure}

\begin{figure}
\begin{center}
\begin{tabular}{c c}
         \includegraphics[width=0.5\textwidth]{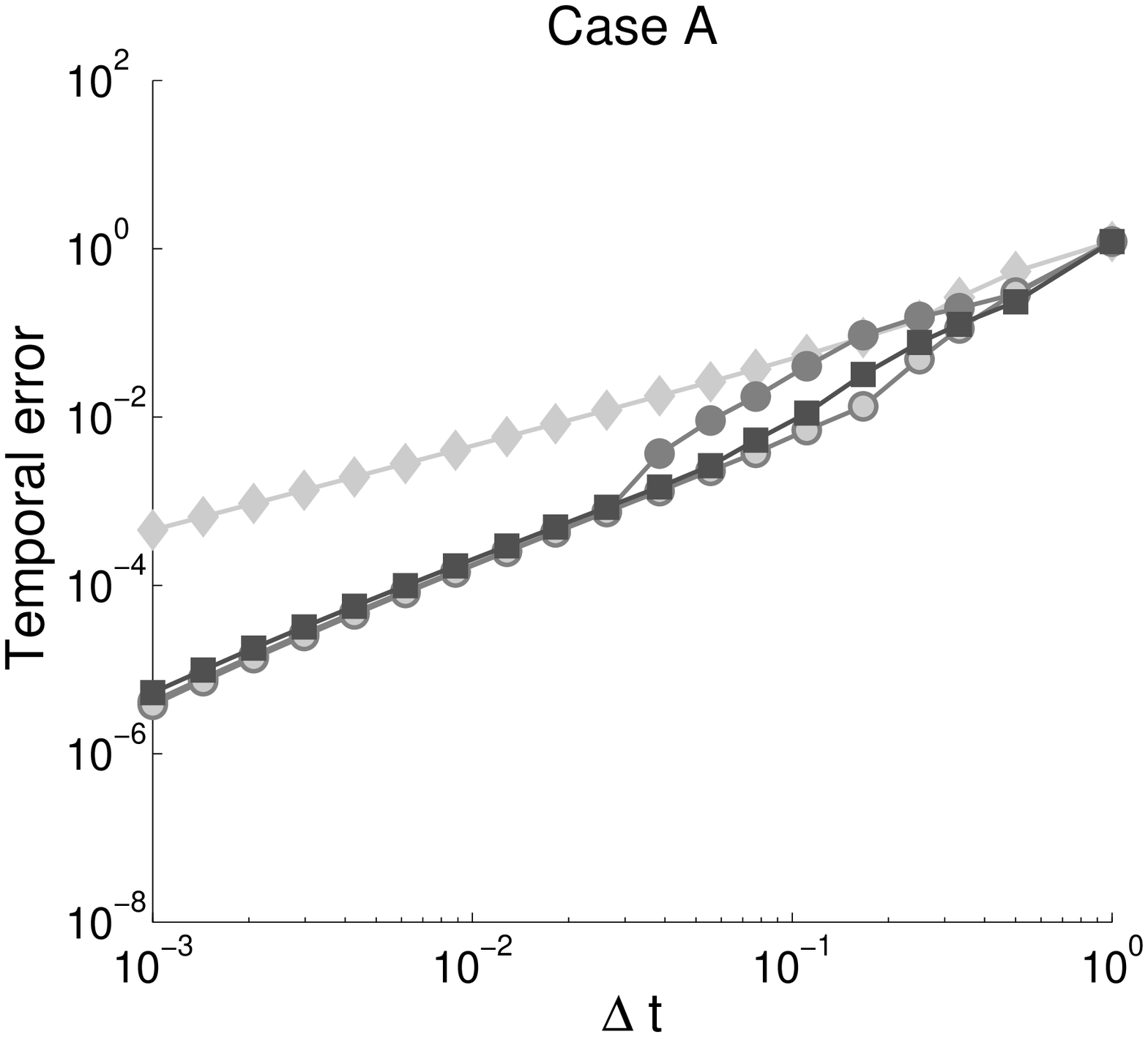}&
         \includegraphics[width=0.5\textwidth]{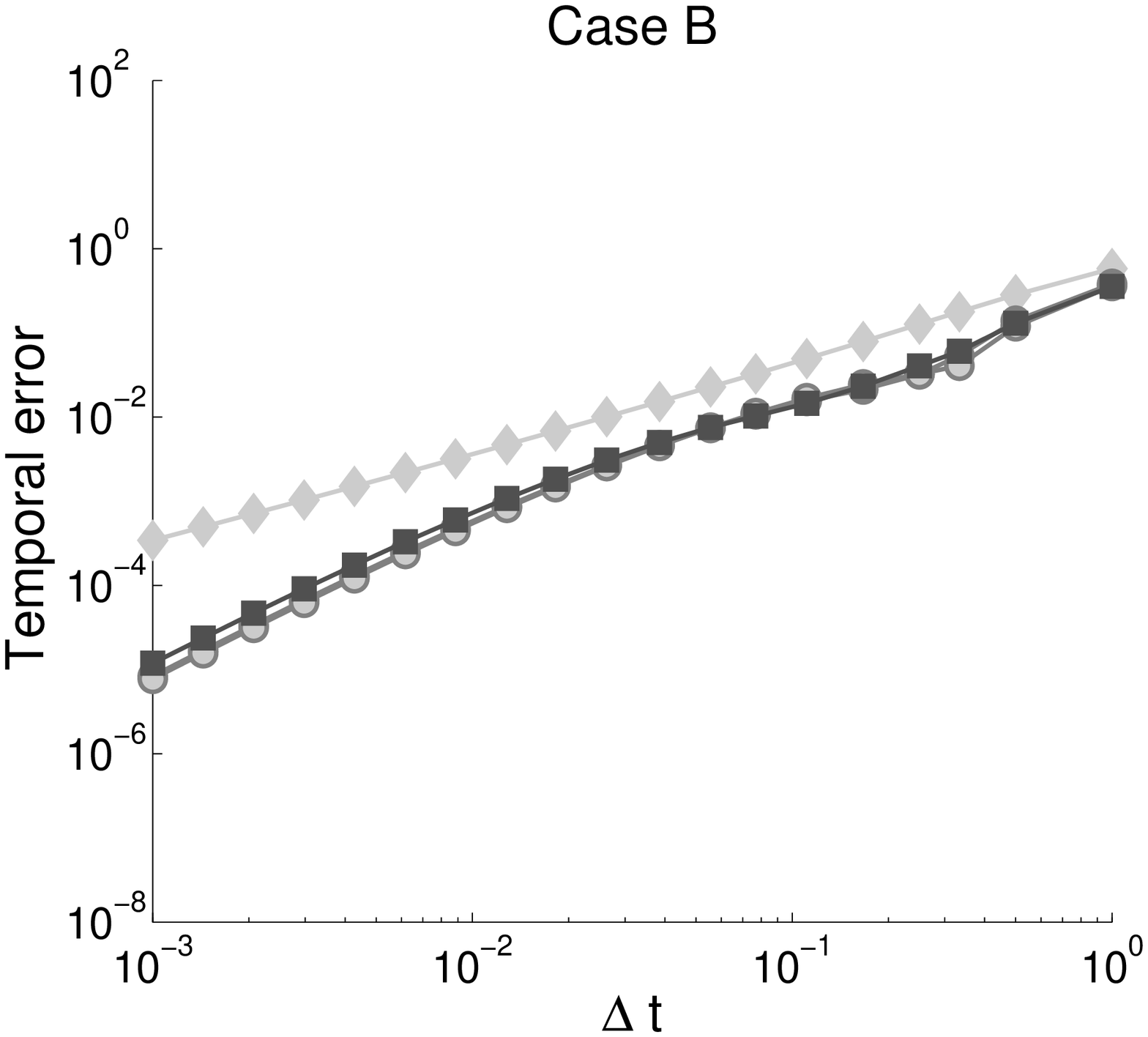}\\
         \includegraphics[width=0.5\textwidth]{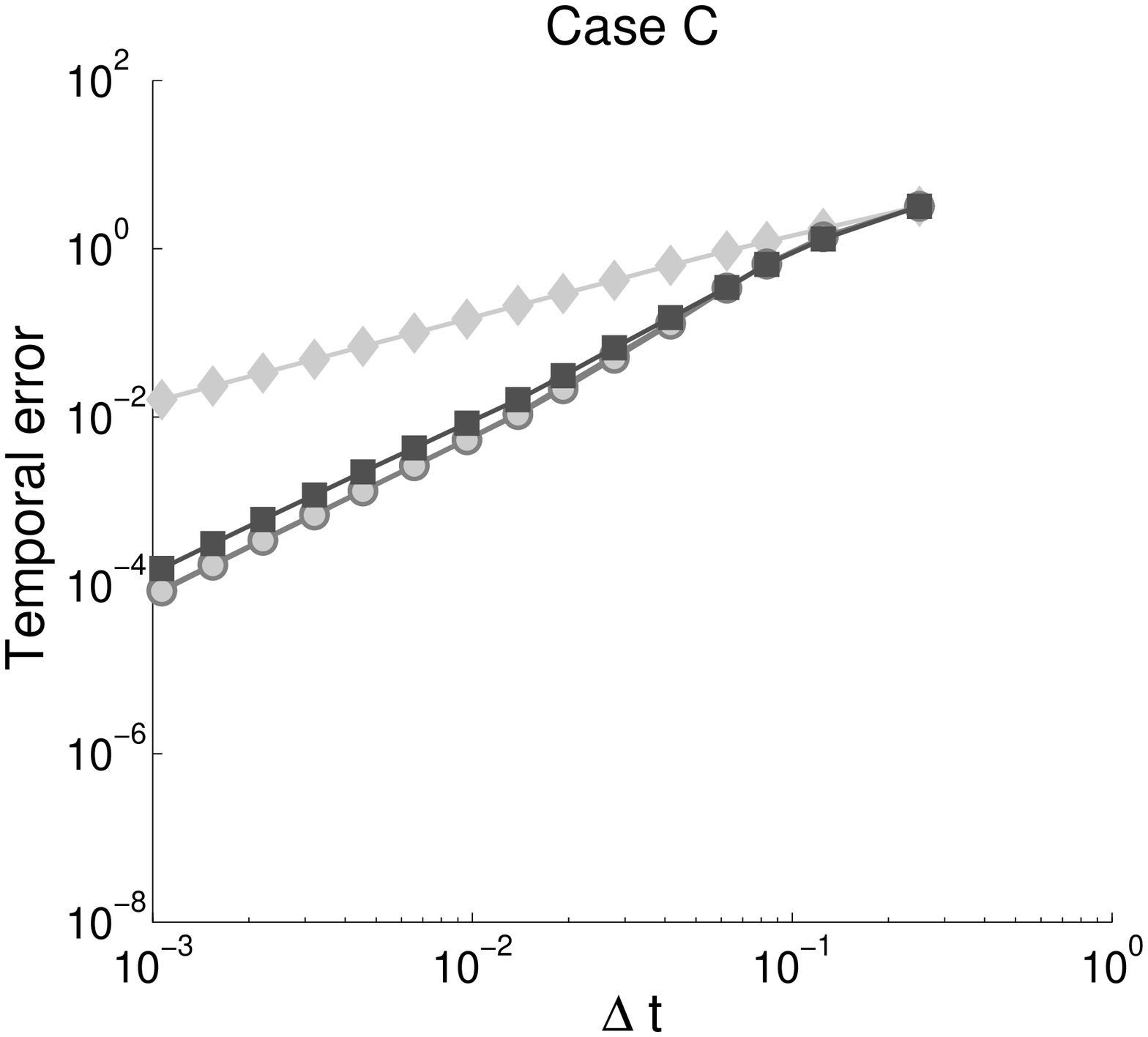}&
         \includegraphics[width=0.5\textwidth]{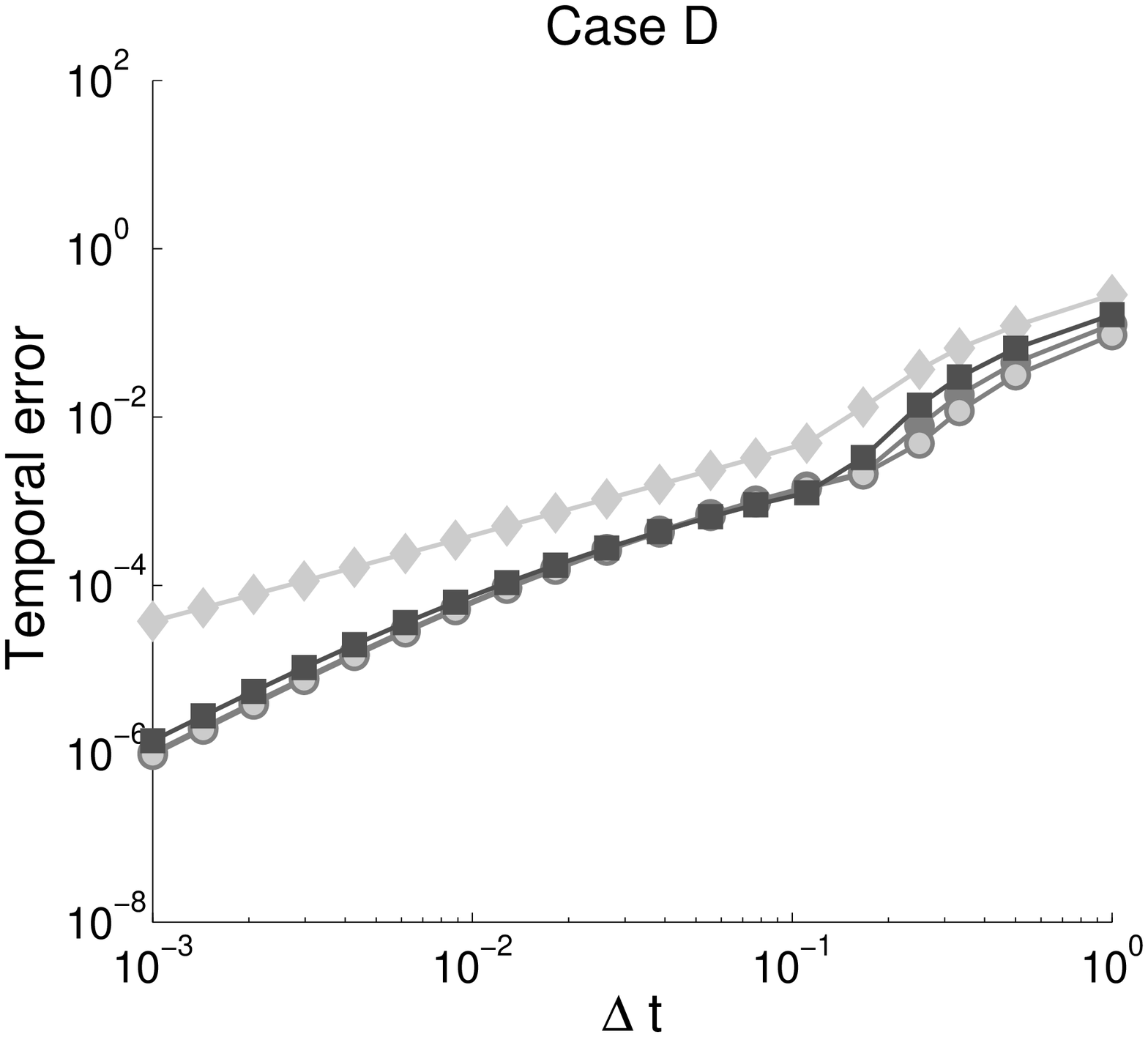}\\
         \includegraphics[width=0.5\textwidth]{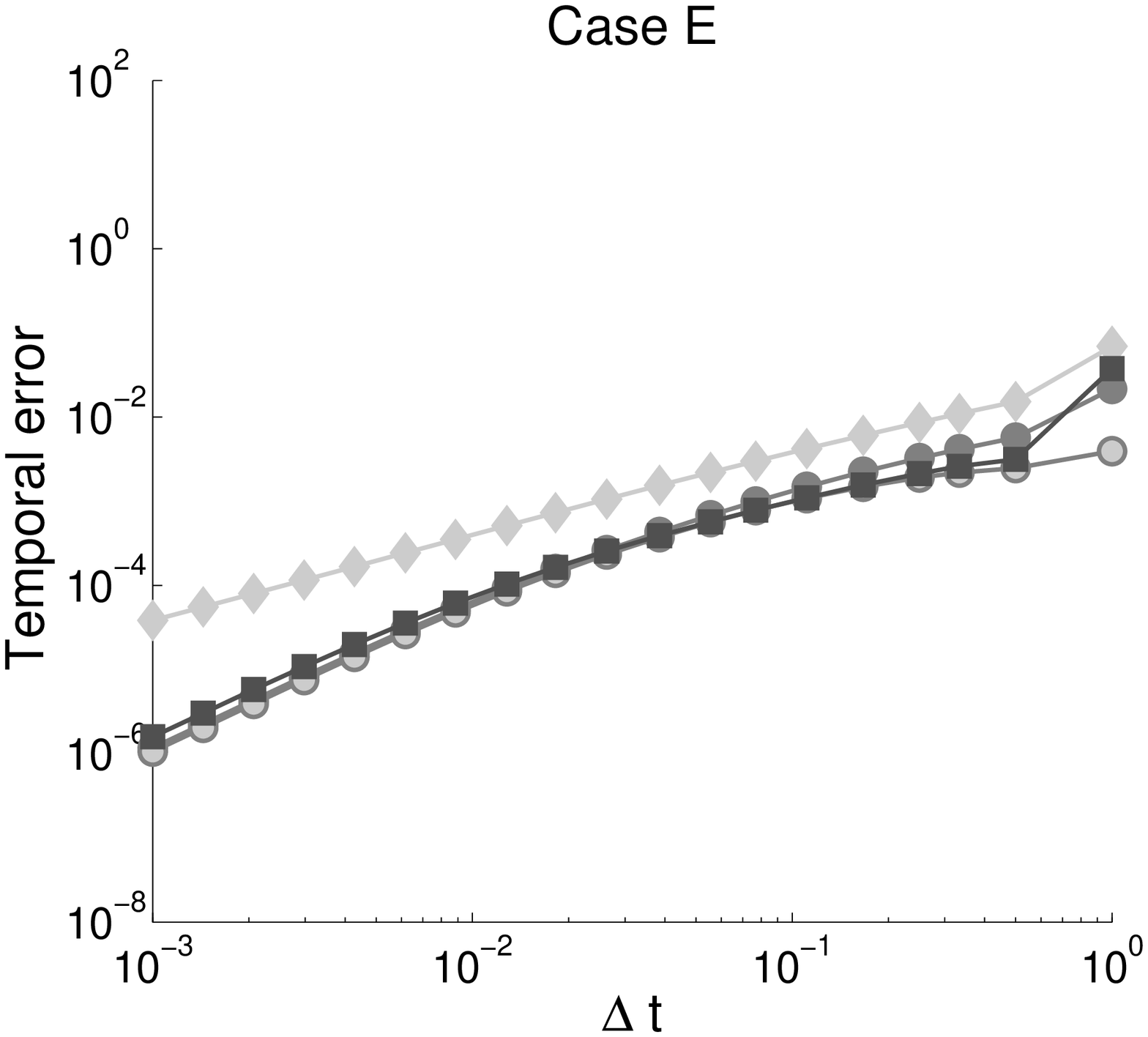}&
         \includegraphics[width=0.5\textwidth]{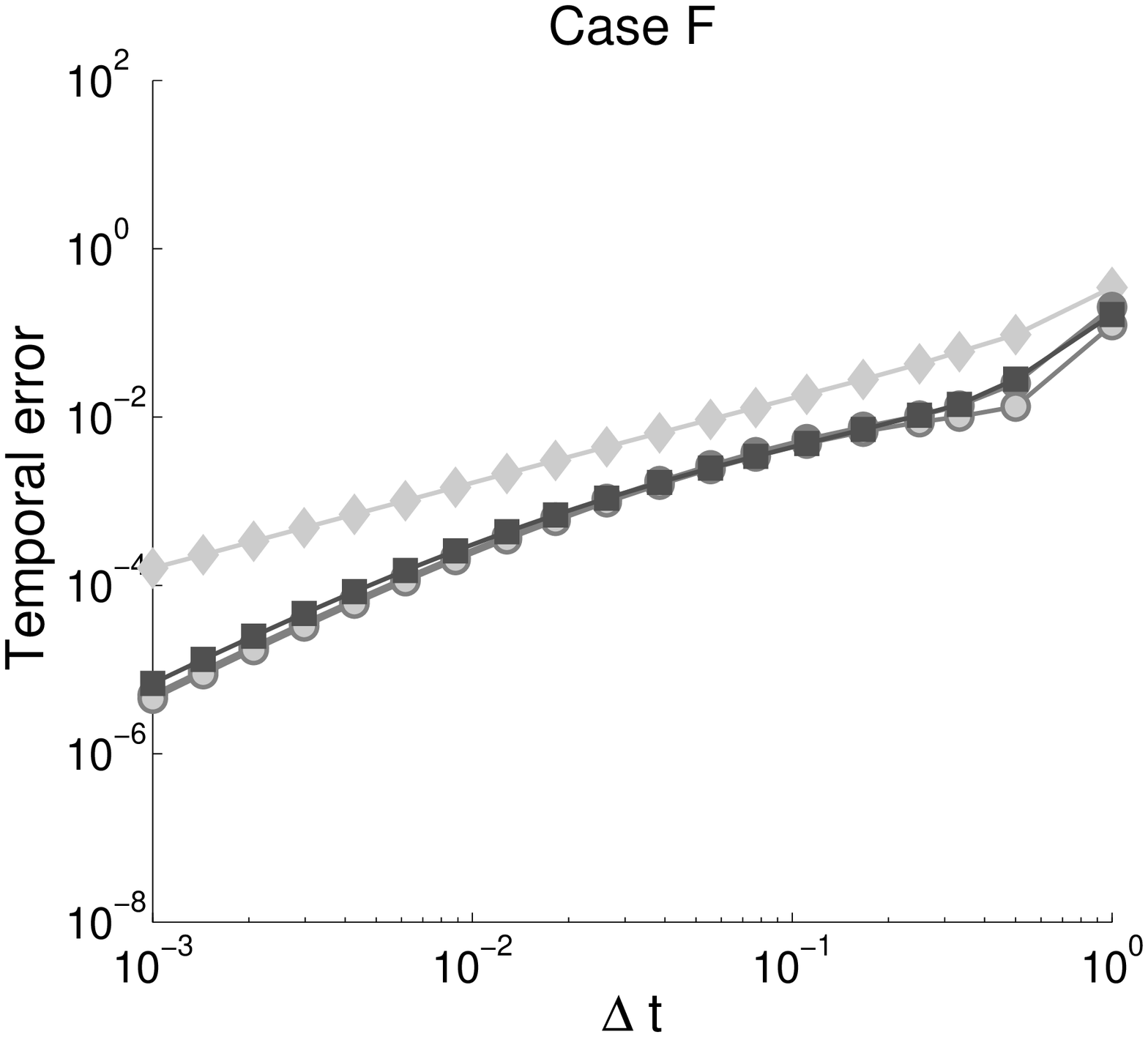}
\end{tabular}
\end{center}
\caption{Temporal discretization errors $\widehat{e}\,(\Delta
t;100,50,50)$ vs $\Delta t$\, for up-and-out call options in all
cases of Table~\ref{cases1} for barrier $B=120$.
ADI schemes: Do with $\theta=\frac{2}{3}$ (diamond), CS with $\theta=\frac{1}{2}$
(dark circle), MCS with $\theta=\max\{\frac{1}{3},\frac{2}{13}(2\gamma+1)\}$
(light circle), and HV with $\theta=\frac{1}{2}+\frac{1}{6}\sqrt{3}$ (square). 
Two initial damping substeps using the Do scheme with $\theta =1$.}
\label{TemporalError2b}
\end{figure}

Figure~\ref{ExactBarrier} displays the numerically obtained up-and-out
call option values in the six cases of Table~\ref{cases1} for barrier
$B=120$ and (sampled) spot interest rates $r \approx 0.02$ on the
$(s,v)$-domain $[0,B]\times [0,1)$.
Here the FD discretization has been applied with $m=50$ and for the 
time discretization the HV scheme is used with $\Delta t = 10^{-2}$.

We study in detail the performance of the four ADI schemes.
Similar to the case of European call options, Figure~\ref{TemporalError2a}
shows the temporal discretization errors $\widehat{e}\,(\Delta t;2m,m,m)$
in the case of up-and-out call options for a sequence of step sizes
$10^{-3} \le \Delta t \le 10^0$ when $m=50$.
As a first observation, it is clear from Figure~\ref{TemporalError2a} that
the unfavorable feature of relatively large temporal errors for moderate
step sizes is more pronounced compared to the case of vanilla options,
especially for the CS scheme, cf Subsection \ref{Eurcall}.
We attribute this to the additional discontinuity of the payoff function
at the barrier $s=B$.
We therefore consider application of a damping procedure at $t=0$.
Instead of performing two substeps at $t=0$ with step size $\Delta t / 2$
by the implicit Euler scheme, which forms a common approach, we employ 
here the Do scheme, with parameter value $\theta=1$.
This is computationally more attractive when dealing with multidimensional 
PDEs.
Figure~\ref{TemporalError2b} shows the temporal discretization errors
in the case of up-and-out call options when two initial substeps with
the Do scheme and $\theta=1$ are applied.
Clearly, the behavior of the temporal error as a function of the step
size has become regular and, in most cases, at only a limited loss of
accuracy for small $\Delta t$ (an exception being the CS scheme in
case E).
Hence, the present damping procedure performs satisfactory.
Applying two substeps of the implicit Euler scheme for the damping would 
yield similar or somewhat smaller temporal errors than those in Figure 
\ref{TemporalError2b}. 
However, we find that this comes at a much higher computational cost, 
also when iterative solvers, like BiCGSTAB, are applied.
We thus infer that damping with the Do scheme is more efficient.

As a main positive conclusion, the numerical results for all ADI schemes
are consistent with an unconditionally stable behavior:
the temporal discretization errors are bounded from above by a moderate
value and decay monotonically as $\Delta t$ decreases, which is obtained
for any value~$m$ tested.
A closer inspection of the results displayed in Figure \ref{TemporalError2b}
yields that the temporal errors behave for sufficiently small $\Delta t$
as $C (\Delta t)^p$ with $p=1.0$ for the Do scheme and $1.6\le p \le 2.0$
for the CS, MCS, HV schemes, with constants $C$.
Experiments with different values of $m$ reveal that both $p$ and $C$
are only weakly dependent on the number of spatial grid points $M$,
indicating that the error behavior is valid in a stiff, hence favorable,
sense.
Note further that the difference in performance between the Do scheme
and the CS, MCS, HV schemes is often less striking than in the case of
vanilla options, but for the latter three schemes combined with damping 
still a higher order and increased accuracy is obtained.

\section{Conclusions and future research}\label{concl}
In this paper we studied ADI schemes in the numerical solution of the
three-dimensional HHW PDE: the Do scheme, the CS scheme, the MCS scheme
and the HV scheme, each with a well~chosen parameter $\theta$.
Extensive experiments have been conducted for six cases of parameter
sets for the HHW model, including correlations that are all nonzero,
time-dependent mean-reversion levels, and short and long maturities.
In three cases the Feller condition is not fulfilled.
We considered both European call options and up-and-out call options.
Our tests have shown that all ADI schemes perform very well in terms
of stability, accuracy and efficiency.
In particular they always reveal an unconditionally stable behavior.
Next, the Do scheme always has a stiff order of convergence equal
to one.
The CS, MCS, HV schemes show a stiff order of convergence equal to
two for European call options and, when combined with damping,
between 1.6 and 2.0 for up-and-out call options.

Based on the numerical experiments and the theoretical stability
results, we find that the MCS scheme with
$\theta = \max\{\tfrac{1}{3},\tfrac{2}{13}(2\gamma+1)\}$
and the HV scheme with
$\theta = \tfrac{1}{2}+\tfrac{1}{6}\sqrt{3}$
are preferable.
Here $\gamma =\max_{ij} |\rho_{ij}|$.
Also the CS scheme with $\theta = \tfrac{1}{2}$ is a good candidate.
For the latter scheme, a damping procedure at $t=0$ is always
recommended.
Damping can be done efficiently, in an ADI fashion, by applying
the Do scheme with $\theta = 1$.

The Do, CS, MCS and HV schemes are expected to perform well and possess
similar favorable properties as obtained in this paper in the numerical
solution of many other three-dimensional PDEs and for other exotic
options.
Also, the ADI schemes can directly be applied, with high efficiency, when
any other FD discretization is employed, as the matrices $A_j(t)$
($1\le j\le 3$) always have a small bandwidth.
We shall investigate other applications in future research.
At the same time, a further theoretical stability analysis of the ADI
schemes will be carried out.


\setcounter{equation}{0}
\section*{Appendix}
Here we give the semi closed-form analytic formula for European call
option values $\varphi(s,v,r,\tau)$ under the HHW model (\ref{SDE})
with $\rho_{13}=\rho_{23}=0$ as derived in Muskulus, In 't Hout,
Bierkens et al (2007).
The notation is adapted to our present situation.
We put $\rho=\rho_{12}$.

The solution presented in loc cit is of a form similar to the
Black--Scholes formula,
\begin{equation*}\label{guess}
\varphi(e^x,v,r,\tau) = e^x P_1(x,v,r,\tau) - K B(r,\tau) P_2(x,v,r,\tau).
\end{equation*}
Here $B(r,\tau)$ denotes the value at time $\tau$ of a zero-coupon bond
that pays 1 at maturity, given that at time $\tau$ the short rate equals
$r$.
For this, it is well-known that
\begin{subeqnarray*}\label{BF}
B(r,\tau) &=& e^{c(r,\tau)} \,, \\
c(r,\tau) &=& -\frac{r}{a}\left(1-e^{-a(T-\tau)}\right)
-\int_\tau^T b(\lambda)\left(1-e^{-a(T-\lambda)}\right)d\lambda
\nonumber\\
          & & +~\frac{\sigma_2^2}{2a^2}\left( T-\tau +
\frac{2}{a}e^{-a(T-\tau)}-\frac{1}{2a} e^{-2a(T-\tau)}-\frac{3}{2a}\right).
\end{subeqnarray*}
The $P_1$, $P_2$ can be viewed as probabilities and are retrieved from
characteristic functions $f_1$, $f_2$ by inversion:
\begin{equation*}\label{INV}
P_j(x,v,r,\tau)=  \frac{1}{2} + \frac{1}{\pi} \int_0^\infty
{\rm Re}\left[\frac{e^{-\imi y\ln K}f_j(x,v,r,\tau;y)}{\imi y}\right]dy
\quad {\rm for}~j=1,2.
\end{equation*}
with $\imi^2 = -1$.
The functions $f_1$, $f_2$ have the form
\begin{eqnarray*}\label{f}
f_1(x,v,r,\tau;y)&=&e^{F_1(\tau;y)+G_1(\tau;y)v+H_1(\tau;y)r+\imi x y},\\
f_2(x,v,r,\tau;y)&=&e^{F_2(\tau;y)+G_2(\tau;y)v+H_2(\tau;y)r+\imi x y-c(r,\tau)}.
\end{eqnarray*}
Let $\delta_1 = 0$, $\delta_2 = 1$ and $j=1,2$.
Then
\begin{equation*}\label{H12}
H_j(\tau;y) = \frac{\imi y -\delta_j}{a} \left( 1-e^{-a(T-\tau)} \right).
\end{equation*}
Next, let
\begin{equation*}
\alpha = \kappa \eta~~,~~\beta_1 = \kappa - \rho\sigma_1~~,~~\beta_2 = \kappa~~,
~~\gamma_1 = \frac{1}{2}~~,~~\gamma_2 = -\frac{1}{2}
\end{equation*}
and
\begin{equation*}
d_j = \sqrt{(\beta_j -\imi \rho\sigma_1 y)^2 -\sigma_1^2(2\imi\gamma_j y -y^2)}~~,~~
g_j = \frac{\beta_j -\imi \rho\sigma_1 y+d_j}{\beta_j -\imi\rho\sigma_1 y -d_j}\,.
\end{equation*}
Then
\begin{equation*}\label{G12}
G_j(\tau;y) = \frac{\beta_j - \imi \rho\sigma_1  y +d_j}{\sigma_1^2}
\left[ \frac{1-e^{d_j (T-\tau)}}{1- g_j e^{d_j (T-\tau)}}\right].
\end{equation*}
Finally,
\begin{eqnarray*}\label{F12}
F_j(\tau;y)
&=& \frac{\alpha}{\sigma_1^2}\left\{(\beta_j - \imi\rho\sigma_1 y +d_j)(T-\tau)
-2\ln \left[ \frac{1- g_j e^{d_j(T-\tau)}}{1-g_j}\right] \right\}
\nonumber \\
&&+ (\imi y-\delta_j)\int_\tau^T b(\lambda)\left(1-e^{-a(T-\lambda)}\right)d\lambda
\nonumber\\
&& + \frac{\sigma_2^2}{2} \left( \frac{\imi y -\delta_j}{a} \right)^2
\left( T-\tau + \frac{2}{a} e^{-a(T-\tau)}-\frac{1}{2a} e^{-2a(T-\tau)}-\frac{3}{2a}\right).
\end{eqnarray*}

The above valuation formula is easily seen to constitute a proper extension
of Heston's (1993) formula, by taking $b(\tau)\equiv r_0$ and $\sigma_2=0$.
It can be approximated to any accuracy, by a direct adaptation of numerical
integration techniques already well studied in the literature for Heston's
formula.
We note that the additional integrals involving the function $b$ can be
exactly determined in our particular case of (\ref{meanr}).

\section*{Acknowledgements}
The authors gratefully acknowledge Peter Forsyth for a stimulating discussion,
convincing them of the proper boundary condition for the HHW PDE at $v=0$ and
pointing them to the work by Ekstr\"{o}m \& Tysk.
The authors also thank Jan Van Casteren for a valuable, unpublished note on
the validity of the HHW PDE.
Furthermore, they are indebted to Sven Foulon for providing an implementation
of Heston's formula for European call options, which we extended to our case.
This work has been supported financially by the Research Foundation -- Flanders,
FWO contract no.~G.0125.08.

\end{document}